\documentclass[pra,twocolumn,showpacs,preprintnumbers,amsmath,amssymb]{revtex4}
\usepackage{graphicx}
\usepackage{dcolumn}
\usepackage{bm}
\usepackage{subfigure}
\usepackage{amssymb}
\usepackage{verbatim}
\usepackage{amscd}
\usepackage{amssymb}
\usepackage{amsmath, amsfonts}
\usepackage{setspace}
\usepackage[]{graphicx}         
\usepackage{amsthm}
\usepackage{natbib}

\theoremstyle{plain}
\newtheorem{theorem}{Theorem}
\theoremstyle{plain}

\theoremstyle{remark}

\theoremstyle{observation}

\theoremstyle{corollary}
\newtheorem{corollary}{Corollary}

\begin{document}


\title{Framework for classifying logical operators in stabilizer codes}


\begin{titlepage}

\author{Beni Yoshida}
\affiliation{Department of Physics, Massachusetts Institute of Technology, Cambridge,
Massachusetts 02139, USA}
\author{Isaac L. Chuang}
\affiliation{Department of Physics, Massachusetts Institute of Technology, Cambridge,
Massachusetts 02139, USA}
\date{\today}

\begin{abstract}
Entanglement, as studied in quantum information science, and non-local
quantum correlations, as studied in condensed matter physics, are
fundamentally akin to each other. However, their relationship is often
hard to quantify due to the lack of a general approach to study both
on the same footing. In particular, while entanglement and non-local
correlations are properties of states, both arise from symmetries of
global operators that commute with the system Hamiltonian. Here, we
introduce a framework for completely classifying the local and
non-local properties of all such global operators, given the
Hamiltonian and a bi-partitioning of the system. This framework is
limited to descriptions based on stabilizer quantum codes, but may be
generalized. We illustrate the use of this framework to study
entanglement and non-local correlations by analyzing global symmetries
in topological order, distribution of entanglement and entanglement
entropy.
\end{abstract}

\maketitle
\end{titlepage}

\section{Introduction \label{introduction}}

In recent years, ideas from quantum information science have become
increasingly useful in condensed matter physics \cite{Lieb,
  Entanglement_Criticality, Entropy_CFT_2D, Entanglement_QPT1, Li08,
  Topo1, Topo2}. In particular, it has been realized that many
interesting physical systems in condensed matter physics may be
described in the language of quantum coding schemes such as the
stabilizer formalism \cite{Toric, MBQC, Graph}.  What is emerging is a
closeness between the two fields, which is a result of the intrinsic
similarity between quantum correlations, as studied in condensed
matter physics, and entanglement, as studied in quantum information
science. The common building block for these studies is the locality
and non-locality of correlations and entanglement in the system.

The notion of non-locality versus locality naturally arises in quantum
information science through the study of quantum entanglement, as
employed in many different constructions.  For example, entangled
states are essential as resources for quantum error-correcting codes,
where encoded logical qubits are typically dispersedly entangled, and
may be transformed among themselves by logical
operators \cite{Nielsen_Chuang}.  In the study of quantum codes,
non-locality arises both in entangled states and in logical operators
which act on entangled states.  Particularly in quantum codes defined
on geometric manifolds, how locally or non-locally logical operators
may be defined is of particular interest since the geometric sizes of
the logical operators determine properties such as the code distance
and rate \cite{Toric}. Also, such geometric sizes are closely related
to the maximum tolerable rate of error in fault-tolerant constructions
using the quantum code, the fault-tolerance threshold
\cite{FT1}. Entangled states can also provide serve as resources for
perfect secret-sharing of classical or quantum information
\cite{secret1}. In such schemes, information is shared between one or
more parties, using a logical operator which is defined jointly over
all the parties in a non-local way.  Again, the key construct is the
logical operator.  Clearly, for quantum codes, secret sharing, and for
other quantum science applications, it is desirable to be able to
determine whether a given local operator can be locally defined inside
some subset of qubits or not.  However, this is often difficult, both
analytically or computationally.  This is a challenge which underlies
broad questions in quantum coding theory, including upper bounds on
code distances of locally defined stabilizer codes \cite{Freedman,
  KC08} and the feasibility of self-correcting memory
\cite{Bacon_Shor, Local_Stabilizer}.

Non-locality and locality is also central to condensed matter physics,
as seen in the study of spin systems, for example. The non-locality of
correlations in such systems has been addressed through entanglement
entropy \cite{Area_Law, Entropy_CFT_2D, Entanglement_Criticality,
  Entanglement_QPT1} and its generalizations \cite{Topo1,Topo2,Li08}
for systems commonly discussed in condensed matter physics.
Renormalization of correlation length scales leads to novel, efficient
numerical simulation schemes using, for example, the matrix product
state \cite{MPS} or tensor product state formalisms
\cite{Entanglement_Renormalization, Gu08}. Much current interest
focuses on topologically ordered states whose correlation extends over
the entire system \cite{Toric, Honeycomb, Z2_insulator,
  String_Net}. Such a global correlation is often a result of global
symmetries of the system Hamiltonian which manifest themselves as the
existence of global operators commuting with the Hamiltonian. These
global operators resulting from the symmetry of the Hamiltonian are
essentially akin to logical operators in quantum codes. Thus, one
might hope that the analysis of logical operators and symmetry
operators could provide a unified approach to understand the
non-locality and locality of correlations and entanglement in
condensed matter physics and quantum coding theory.

However, the connection between locality and non-locality in quantum
coding theory and condensed matter physics is currently incomplete. In
particular, the relation between the quantum entanglement discussed in
quantum coding theory and the correlations studied in condensed matter
physics has not been fully understood. A general framework to study
the non-local and local properties of such systems will be a necessary
first step to clarify the relation between two fields. Such a
framework should be applicable to many classes of systems described
through quantum coding schemes; and it should enable quantification of
the degree of non-locality of correlations in any bi-partition or
multi-partition of the system. Also, since many strongly entangled
systems possess a degenerate ground state, the framework should study
not only single states, but also the entire ground state
space. Finally and most importantly, the framework would need to
provide a systematic procedure to obtain all the logical operators in
a computationally tractable way. The framework should classify all the
logical operators according to their localities and non-localities so
that the coding properties and their relation with the non-local
correlation of the system can be studied on the same footing.

There are several pioneering works which have taken first steps toward
such a framework. A first analysis leading in this direction was
conducted for graph states, which are multi-partite entangled states
corresponding to mathematical graphs \cite{Graph}. It was found that
multi-partite entanglement could be characterized and quantified
completely in terms of the Schmidt measure.  Later, the problem of
finding a systematic framework was further investigated through the
study of stabilizer states \cite{FC04}. A complete characterization
and quantification of the degree of non-locality of correlations in
any bi-partite stabilizer state were provided. A generalization to
multi-partite entanglement was also discussed in the same
work. Recently, another important advance was made by extending this
work to stabilizer codes in the context of two party quantum
information encoding \cite{Wilde09}. These works have initiated
the investigation of non-local correlations of the systems described
through quantum coding schemes, but largely only for individual
states. The need is to generalize these approaches to an analysis of
spaces of degenerate ground states, in order to connect quantum coding
theory to condensed matter physics.
 
Here, we present a new framework which builds on the current
techniques \cite{Graph, FC04} and reveals the non-local correlations
in stabilizer codes through the analysis of logical operators in a
bi-partition. In particular, the framework classifies all the logical
operators according to their non-localities and localities, and
computes all the logical operators along with the classification. The
framework begins by identifying a group of operators, \textit{the
  overlapping operator group}, which is at the heart of the
correlations over two separated subsets of qubits. A new theoretical
tool, a \textit{canonical representation}, is included in the
framework to analyze the overlapping operator group and to obtain
logical operators along with their non-localities. This framework is
specifically limited to stabilizer codes with a bi-partition into two
complementary subsets.
 
The presentation of this framework is organized as follows. In
section~\ref{sec:review}, we give a brief review of the stabilizer
formalism. Then, in section~\ref{sec:setups}, we introduce a
classification of logical operators based on their localities and
non-localities. We define the overlapping operator group and describe
the theoretical tool to analyze it. In section~\ref{sec:logical}, we
provide a systematic procedure to compute all the logical operators
based on the classification. And in section~\ref{sec:discussion}, we
illustrate the use of our framework to study the non-local property of
the stabilizer formalism by showing three specific examples,
considering the global symmetries in topological order, the
distribution of multi-partite entanglement, and entanglement entropy.

\section{Review of Stabilizer Codes }
\label{sec:review}

We begin with a review of stabilizer codes and their logical operators
\cite{Stab1, Stab2}. Consider an $N$ qubit system governed by the Hamiltonian
\begin{equation}
H_{stab} = - \sum_{i} S_{i} \label{stabHam}
\end{equation}
where the interaction terms $S_{i}$ are inside the \textit{Pauli operator
group}
\begin{align}
\mathcal{P} = \langle iI, X_{1},Z_{1},\dots , X_{N},Z_{N} \rangle
\end{align} 
which is generated from local Pauli operators $X_{i}$ and $Z_{i}$ acting on
each of the $N$ single qubits. In \textit{stabilizer codes}, interaction terms
$S_{i}$ are called \textit{stabilizer generators} and they commute with each
other, obeying $[S_{i},S_{j}]=0$ for $\forall i, j$. The \textit{stabilizer
group} 
\begin{align}
\mathcal{S} = \langle \{ S_{i}, \hspace{1ex} \forall i \} \rangle,
\end{align}
which is generated from all the stabilizer generators $S_{i}$, is a
self-adjoint Abelian subgroup of the Pauli operator group which does
not contain $- I$. Operators inside the stabilizer group $\mathcal{S}$
are called \textit{stabilizers}. This \textit{stabilizer Hamiltonian}
is exactly solvable since stabilizer generators $S_{i}$ commute each
other and $S_{i}^{2}=I$. The entire Hilbert space can be decomposed
into a direct sum of subspaces with respect to $s_{i}=\pm 1$, where
$s_{i}$ are the eigenvalues of the stabilizer generators $S_{i}$, as
\begin{align}
\mathcal{H} = \bigoplus_{\vec{s}} \mathcal{H}_{\vec{s}}.
\end{align}
Each of decomposed subspaces $\mathcal{H}_{\vec{s}}$ is characterized
by eigenvalues $\vec{s}=(s_{1}, s_{2}, \cdots )$. The ground state
space is $\mathcal{H}_{\vec{1}} \equiv \mathcal{H}_{1,1,\cdots}$ where
each ground state $|\psi\rangle$ is \textit{stabilized} as
$S_{i}|\psi\rangle = |\psi \rangle$ for $\forall i$ since the choice
of $\vec{s}=\vec{1}$ minimizes the energy of the stabilizer
Hamiltonian $H_{stab}$. Now let us denote the number of nontrivial
generators for a group of operators $\mathcal{O}$ as
$G(\mathcal{O})$. If the number of generators for $\mathcal{S}$ is
$G(\mathcal{S})=N-k$, we notice that there are $2^{k}$ ground states
for this stabilizer code. Then, quantum information can be stored
among degenerate ground states inside $\mathcal{H}_{\vec{1}}$ by
assigning eigenstates of qubits $|\tilde{0}\rangle$ and
$|\tilde{1}\rangle$ to each pair of degenerate ground states. The
encoded qubits in the ground state space are called \textit{logical
  qubits}.

Logical qubits and their coding properties can be characterized by
\textit{logical operators}. Logical operators are defined as Pauli operators
$\ell \in \mathcal{P}$ which commute with the entire Hamiltonian,
$[\ell,H_{stab}]=0$, but $\ell \not\in \langle \mathcal{S}, iI \rangle$. Here, $iI$
is inserted since $\mathcal{P}$ includes $iI$ while $\mathcal{S}$ does not.
Since logical operators commute with the stabilizer Hamiltonian $H_{stab}$,
when a logical operator $\ell$ is applied to one of the ground states
$|\psi_{0}\rangle$ of the Hamiltonian, the resulting state $\ell|\psi_{0}\rangle$
is also a ground state of the Hamiltonian. Therefore, logical operators can
transform the ground states among themselves and manipulate logical qubits
encoded inside the ground state space $\mathcal{H}_{\vec{1}}$. One can also
notice that the applications of stabilizers keep the ground state unchanged
since they do not change the eigenvalues $\vec{s}$ of stabilizer generators
$S_{i}$. Then, stabilizers inside $\mathcal{S}$ can be viewed as trivial
logical operators. Since the application of stabilizers does not change the
properties of logical operators which act on logical qubits, two logical
operators $\ell$ and $\ell'$ are called \textit{equivalent} when $\ell \ell' \in \langle
\mathcal{S}, iI \rangle$. The equivalence between two logical operators $\ell$
and $\ell'$ are represented as $\ell \sim \ell'$.

In order to obtain all the logical operators which are not equivalent each
other, it is convenient to consider the \textit{centralizer group} 
\begin{align}
\mathcal{C} = \langle \{ U \in \mathcal{P} \mid [U,S_{i}] =0,\hspace{1ex} \forall i \} \rangle
\end{align}
and its quotient group with respect to $\mathcal{S}$
\begin{align}
\mathcal{C} / \mathcal{S} = \langle i \mathcal{S}, \ell_{1} \mathcal{S} , r_{1} \mathcal{S}, \dots , \ell_{k} \mathcal{S} , r_{k} \mathcal{S} \rangle.
\end{align}
The centralizer group $\mathcal{C}$ includes all the Pauli operators which
commute with any stabilizer generators $S_{i}$. All the stabilizer generators
and logical operators are inside $\mathcal{C}$ from their definitions. The
$2k$ nontrivial representatives $\ell_{i}$ and $r_{i}$ ($i=1, \cdots
,k$) for $\mathcal{C} / \mathcal{S}$ are \textit{independent} logical
operators which commute with each other, except for $\{ \ell_{i},
r_{i} \} = 0$. Choices of $2k$ representatives are not unique (see
discussion in section~\ref{sec:3C}).

Each of the $2k$ logical operators can decompose the ground state space into a
direct product of $k$ subsystems
\begin{equation}
|\psi\rangle = \bigotimes_{i=1}^{k} (\alpha_{i} |\tilde{0} \rangle_{i} + \beta_{i} |\tilde{1} \rangle_{i}) 
\end{equation}
with 
\begin{equation}
\begin{split}
\ell_{i} |\tilde{0}\rangle_{i}  &= |\tilde{0}\rangle_{i}\\
\ell_{i} |\tilde{1}\rangle_{i}  &= -|\tilde{1}\rangle_{i} \\
r_{i} |\tilde{0}\rangle_{i}  &= |\tilde{1}\rangle_{i} \\
r_{i} |\tilde{1}\rangle_{i}  &= |\tilde{0}\rangle_{i}
\end{split}
\end{equation}
where $| \tilde{0} \rangle_{i}$ and $|\tilde{1} \rangle_{i}$ represent a
subsystem which supports a logical qubit characterized by the $i$-th
anti-commuting pair of logical operators $\ell_{i}$ and $r_{i}$. Therefore, $\ell_{i}$ and $r_{i}$ can be viewed as Pauli operators $Z$ and $X$ acting on logical qubits spaned by $| \tilde{0} \rangle_{i}$ and $|\tilde{1} \rangle_{i}$.

The robustness of the quantum code can be measured by how far two
encoded states are apart, which is quantified through the notion of
\textit{code distances}. Since logical operators transform encoded
states among themselves, the existence of logical operators supported
by a large number of qubits may imply that encoded states are far
apart. The code distance of a stabilizer code is defined through the
sizes of logical operators as follows:
\begin{align}
d = \min_{U \in \mathcal{C}, \not\in \mathcal{S} } w(U) 
\end{align}
where $w(U)$ is the number of non-trivial Pauli operators in $U \in
\mathcal{P}$. Essentially, $w(U)$ measures the \textit{size} of Pauli
operator $U$. The code distance $d$ is the smallest possible size of
all the logical operators in a stabilizer code. One of the ultimate
goals in quantum coding theory is to achieve a larger code distance
$d$ for a fixed $N$ in the presence of some physical constraints on
stabilizer generators $S_{i}$, such as locality and translation
symmetries of the stabilizer generators \cite{Local_Stabilizer}.

\section{The Framework \romannumeral 1}
\label{sec:setups}

Now let us return to the central problem of this paper, concerning
locality and non-locality in the stabilizer formalism: given a
stabilizer code $\mathcal{S}$ which is split into two subsets of
qubits $A$ and $B= \bar{A}$, the problem is to unveil the local and
non-local elements in the stabilizer group $\mathcal{S}$ and logical
operators $\ell_{i}$ and $r_{i}$. By \textit{local}, we mean
that a given operator be represented only through qubits inside one of
the two subsets, $A$ or $B$. By \textit{non-local}, we mean that a
given operator cannot be defined either inside $A$ or $B$ alone.

The bi-partite entanglement of a stabilizer state, which is a subclass
of stabilizer codes with $k=0$, has been studied \cite{FC04} by
analyzing the local and non-local elements in the stabilizer group
$\mathcal{S}$. However, the bi-partite properties of logical operators
(versus states) are what interest us, as they are essential to the
study of systems with degenerate ground states. The non-locality and
locality of logical operators are crucial factors which determine the
coding properties of quantum codes. Also, since logical operators may
serve as indicators of the non-local correlations of the system, the
study of the non-locality of logical operators addresses intrinsic
relationships between condensed matter physics and quantum coding
theory.  Thus, we attempt to provide some unity to approaching the
needs of both fields, by broadening prior studies to encompass the
locality and non-locality of logical operators as well as logical
states.

\begin{figure}[ht!]
\begin{center}
\includegraphics[width=1.0\linewidth]{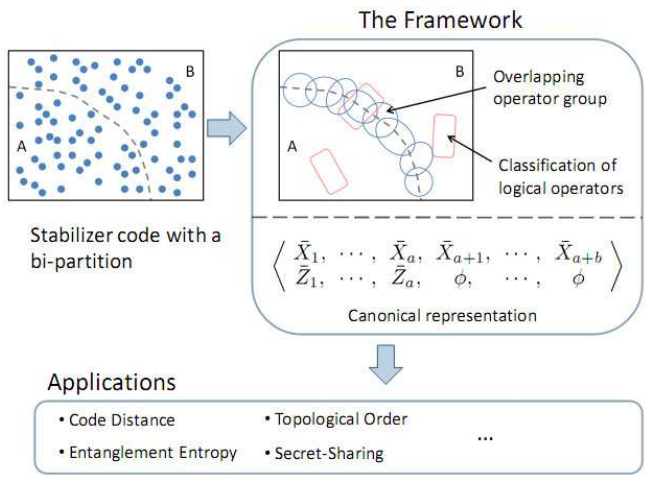}
\caption{The framework to study the non-local properties of stabilizer
  codes. The input to the framework is a stabilizer code with a
  bi-partition. The output of the framework is non-local properties of
  the code. On the left hand side of the figure, a stabilizer code
  with a bi-partition is described. The dots represents qubits in the
  system while the dotted line represents a bi-partition of qubits
  into two subsets $A$ and $B$. The framework consists of the three
  elements. First, the logical operators are classified based on their
  localities and non-localities. Here, the rectangles (red online)
  represent logical operators. Then, the overlapping operator group is
  generated from the overlap of stabilizer generators $S_{i}$. The
  circles (blue online) represent stabilizer generators at the
  boundary between $A$ and $B$. Finally, the framework analyzes the
  overlapping operator group through the canonical representation.  }
\label{fig:tactics}
\end{center}
\end{figure}

In this section and the next, we describe our approach by presenting a
framework which provides a systematic and efficient method to classify
logical operators according to their locality and non-locality in a
bi-partition of stabilizer codes. This framework also obtains all the
logical operators along with their classifications.  The framework
consists of three steps, depicted in Fig.\ref{fig:tactics}.

First, the framework classifies all the $2k$ logical operators by
whether or not they have equivalent logical operators which can be
defined separately inside two complementary subsets $A$ or $B$. This
classification is useful when coding properties of stabilizer codes
are discussed. Also, with the help of the classification of logical
operators, the analysis on logical qubits, which are described by a
pair of anti-commuting logical operators, are simplified.

Second, the framework identifies a group of operators that arises
naturally from the overlap of stabilizer generators at the boundary
between $A$ and $B$. We call this group of operators the
\textit{overlapping operator group}. The overlapping operator group
includes all the necessary clues to study the correlations over $A$
and $B$ in a bi-partition and plays a central role in the study of
logical operators in a bi-partition. However, this group contains
elements that do not commute with each other. This fact makes the
analysis of the overlapping operator group difficult.

Third, to deal with the challenge of analyzing the overlapping
operator group, the framework introduces a new theoretical tool, a
canonical representation which analyzes the overlapping operator
group.  The framework analyzes the overlapping operator group with a
canonical representation and gives a method to compute all the logical
operators in a computationally tractable way.  Also, with the help of
a canonical representation, all the logical operators obtained can be
easily and systematically classified based on their localities and
non-localities.

Below, we first introduce the classification of logical operators in
section \ref{sec:3A}.  Then, we define the overlapping operator group in
section \ref{sec:3B} and describe the canonical representation in section
\ref{sec:3C}.  We also present a systematic method to obtain the canonical
representation, not only for the overlapping operator group, but also
for any subgroup of the Pauli operator group $\mathcal{P}$, in section
\ref{sec:3D}.  Later, we describe a method to compute logical operators along
with the classification, in section \ref{sec:logical}.

\subsection{Classification of Logical Operators}
\label{sec:3A}

All the logical operators may be classified based on their localities
and non-localities. The framework accomplishes this in the following
way.  First, the set of logical operators which can be supported only
with qubits inside a subset $A$ is defined as $L_{A}$. Let us denote
the projection of an operator $O$ onto $A$ as $O|_{A}$. This keeps
only the non-trivial Pauli operators which are inside $A$ and
truncates Pauli operators acting outside the subset $A$. Using this
notation, $L_{A}$ may be represented as
\begin{align}
L_{A} = \{ \ell \in \mathcal{C} | \exists U \in \mathcal{S}, \hspace{1ex} (U\ell)|_{B}=I \}.
\end{align}
$L_{A}$ includes all the logical operators which can be
\textit{shrunk} into a subset $A$ by applying an appropriate
stabilizer generator. $L_{A}$ also includes the stabilizer generators
defined inside $A$, which may be viewed as trivial logical
operators. Note that $L_{A}$ forms a group. When the projections
$(U\ell)|_{B}$ are considered, the arbitrariness resulting from trivial
phase $iI$ in the Pauli operator group $\mathcal{P}$ is neglected
since it does not affect the properties of logical operators. We also
define the set of logical operators which can be supported with qubits
inside a complementary subset $B=\bar{A}$ as $L_{B}$ in a way similar
to $L_{A}$. Since logical operators inside $L_{A}$ or $L_{B}$ can be
defined inside localized subsets in a bi-partition, we call them
\textit{localized logical operators}.

We may further classify localized logical operators in $L_{A}$ by
considering whether they can be also defined inside B or not. Let
$M_{AB}$ be the set of all the logical operators which can be defined
both inside $A$ and inside $B$, represented as
\begin{align}
M_{AB} &= \{ \ell \in \mathcal{C} | \ell \in L_{A}, L_{B} \}.
\end{align}
Note that $M_{AB}$ also forms a group. Next, let $M_{A}$ be the set of
all the logical operators which can be defined inside $A$, but cannot
be defined inside $B$ and define $M_{B}$ similarly. $M_{A}$ and
$M_{B}$ may be represented as
\begin{equation}
\begin{split}
M_{A} &= \{ \ell \in \mathcal{C} | \ell \in L_{A},\not \in  L_{B} \}\\
M_{B} &= \{ \ell \in \mathcal{C} | \ell \not \in  L_{A}, \in L_{B} \}.
\end{split}
\end{equation}
Finally, let $M_{\phi}$ be the set of all the logical operators which
cannot be defined either inside $A$ or $B$, represented as
\begin{align}
M_{\phi} = \{ \ell \in \mathcal{C} | \ell \not \in L_{A},L_{B}  \}.
\end{align}

\begin{figure}[ht!]
\begin{center}
\includegraphics[width=1.0\linewidth]{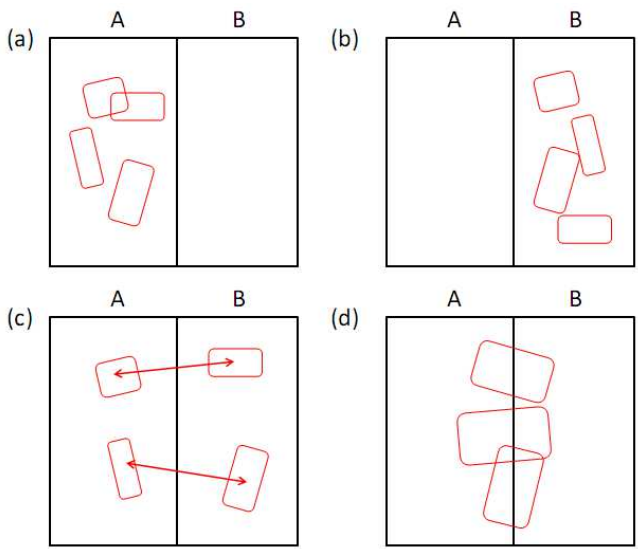}
\caption{The classification of logical operators. 
Each rectangle represents logical operators. 
(a) $M_{A}$. This subset includes logical operators defined only inside $A$. 
(b) $M_{B}$. 
(c) $M_{AB}$. Two logical operators connected through a red arrow are equivalent, each defined inside $A$ or $B$. 
(d) $M_{\phi}$. Non-local logical operators are defined jointly over $A$ and $B$. }
\label{classification}
\end{center}
\end{figure}

We explicitly show the classifications into four sets of logical
operators in Fig.\ref{classification}. Logical operators in $M_{AB}$
have two equivalent representations which can be defined only inside
$A$ or only inside $B$ respectively. On the other hand, logical
operators in $M_{\phi}$ always have support both inside $A$ and $B$ in
a non-local way. We call them \textit{non-local logical operators}
since which are defined over $A$ and $B$ jointly. $L_{A}$ and $L_{B}$
are related to $M_{A}$, $M_{B}$, $M_{AB}$ and $M_{\phi}$ as follows:
\begin{equation}
\begin{split}
L_{A} &= M_{A} \cup M_{AB} \\ L_{B} &= M_{B} \cup M_{AB}
\\ \mathcal{C} &= M_{A} \cup M_{B} \cup M_{AB} \cup M_{\phi}.
\end{split}
\end{equation}

When considering the coding properties of a stabilizer code, the
localized logical operators inside $L_{A}$ and $L_{B}$ become
important, since the code distance $d$ can be upper bounded by the
number of qubits inside each of complementary subsets $A$ and $B$. On
the other hand, non-local logical operators play an important role in
the non-local correlations and entanglement over $A$ and $B$ in the
ground state space. Therefore, \textit{the classification of logical
  operators can serve as a guide to study the non-local properties of
  stabilizer codes}.

\subsection{Overlapping Operator Group}
\label{sec:3B}

Though the classification based on the localities and non-localities
of logical operators are important concepts for describing the
non-locality of the system, it is generally difficult to actually
compute explicit logical operators and to computationally study their
localities and non-localities. For example, if we want to see whether
a given logical operator can be defined inside a localized subset $A$
or not, using the most obvious ``exhaustive elimination'' approach, we
would need to check whether or not there exists a stabilizer operator
which can shrink the corresponding logical operator inside $A$, one by
one. However, such an approach would not allow us to study the
non-local properties of stabilizer codes systematically, since we
would need to repeat the whole computation for each of logical
operators. Clearly, such an direct approach is computationally
intractable.  Instead, what is needed is a more systematic procedure,
which allows efficient computation of all the logical operators, in
order to setup a general framework. In this subsection, we give a
definition of the \textit{overlapping operator group}, which will
become essential in the computation of logical operators based on the
classification of section \ref{sec:logical}.

The group we are interested in follows directly from considering a
general property of localized logical operators inside $A$. Logical
operators $\ell$ defined inside a subset $A$ must commute with all the
stabilizer generators $S_{i}$ which have overlap with qubits inside
$A$. So $\ell$ has the property that $\ell|_{B}=I$ with $[\ell , S_{i}|_{A}]=0$
for $\forall i$.  We define the group of operators generated from
$S_{i}|_{A}$ as the \textit{overlapping operator group},
\begin{align}
\mathcal{O}^{A} = \langle  \{ S_{i}|_{A},\hspace{1ex} \forall i \} \rangle.
\end{align}
Since localized logical operators in $A$ can be computed so that they
commute with all the operators in $\mathcal{O}^{A}$, the analysis of
$\mathcal{O}^{A}$ is at the heart of computing localized logical
operators.

\begin{figure}[ht!]
\begin{center}
\includegraphics[width=0.9\linewidth]{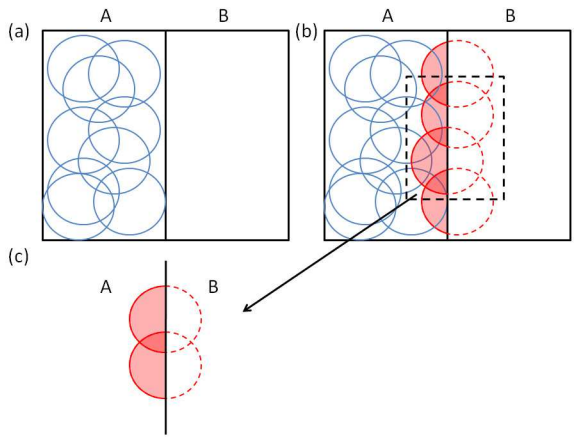}
\caption{An illustration of the differences between the projection
  (restriction) $\mathcal{S}_{A}$ and the overlap
  $\mathcal{O}^{A}$. Each circle represents the stabilizer generators
  supported by qubits inside the circle. (a) The restriction
  $\mathcal{S}_{A}$ generated from all the stabilizer generators
  defined inside $A$. (b) The overlapping operator group
  $\mathcal{O}^{A}$ generated from all the overlaps of stabilizer
  generators with $A$. $\mathcal{O}^{A}$ includes left parts (shaded
  regions) of the stabilizer generators at the boundary between $A$
  and $B$ in addition to stabilizer generators inside $A$. (c) Two
  stabilizer generators at the boundary between $A$ and $B$. Though
  stabilizer generators commute with each other, their projections
  onto $A$ do not necessarily commute with each other. }
\label{fig1}
\end{center}
\end{figure}

Here, we emphasize the difference between the \textit{restriction and
  overlap} of the stabilizer group (Fig.\ref{fig1}). We define the
\textit{restriction} of a group of operators $\mathcal{G}$ into $A$ as
\begin{align}
\mathcal{G}|_{A} = \langle \{ U \in  \mathcal{G} \mid U|_{\bar{A}} =I  \} \rangle.
\end{align}
Therefore, $\mathcal{G|}_{A}$ contains all the operators in
$\mathcal{G}$ which are defined inside $A$. In this paper, we discuss
$\mathcal{S}_{A}\equiv \mathcal{S|}_{A}$, $\mathcal{C}_{A}\equiv
\mathcal{C|}_{A}$ and $\mathcal{P}_{A} \equiv \mathcal{P|}_{A}$ which
are restrictions of the stabilizer group $\mathcal{S}$, the
centralizer group $\mathcal{C}$ and the Pauli operator group
$\mathcal{P}$.

In contrast to the restriction $\mathcal{S}_{A}$, the projected
stabilizer generators $S_{i}|_{A}$ do not necessarily commute with
each other. To see this explicitly, let us pick out two stabilizer
generators $S_{1}$ and $S_{2}$ which are defined at the boundary
between two subsets $A$ and $B$ (Fig.\ref{fig1}(c)), where $B$ is the
complement of $A$.  Represent the projections of $S_{1}$ and $S_{2}$
as
\begin{equation}
\begin{split}
S_{1}|_{A} &= U_{1} \\
S_{1}|_{B} &= V_{1} \\
S_{2}|_{A} &= U_{2} \\
S_{2}|_{B} &= V_{2} .
\end{split}
\end{equation}
Note that $U_{1},U_{2} \in \mathcal{O}^{A}$ since $U_{1}$ and $U_{2}$
are the projections of stabilizer generators onto $A$. Since
$[S_{1},S_{2}]=0$, we have $[U_{1} V_{1} , U_{2} V_{2} ]=0$. This
leads to either $[U_{1},U_{2}]=0$ and $[ V_{1}, V_{2}]=0$ or $\{
U_{1}, U_{2} \}=0$ and $\{ V_{1}, V_{2} \}=0$. Thus, $U_{1}$ and
$U_{2}$ may anti-commute with each other and the overlapping operator
group $\mathcal{O}^{A}$ can be a group of anti-commuting Pauli
operators.

\subsection{Canonical Representation}
\label{sec:3C}

Since the overlapping operator group $\mathcal{O}^{A}$ may include
elements which anti-commute with each other, the analysis of
$\mathcal{O}^{A}$ becomes more complicated than the analysis of the
original stabilizer group $\mathcal{S}$. Here, we introduce a new
approach which we call the \textit{canonical representation} of an
operator group. The canonical representation captures a concept which
naturally arises in order to extract physical properties of logical
operators from a group of anti-commuting Pauli operators. This
representation leads to the computation of all the logical operators
in each of four sets $M_{A}$, $M_{B}$, $M_{AB}$ and $M_{\phi}$ as we
shall see in the next section.

Let us first give the definition of the canonical representation for
the overlapping operator group $\mathcal{O}^{A}$. In the canonical
representation, $\mathcal{O}^{A}$ can be represented as
\begin{align}
\mathcal{O}^{A} = \langle iI , \bar{X}_{1} , \dots , \bar{X}_{a+b}, \bar{Z}_{1} , \dots , \bar{Z}_{a} \rangle
\end{align}
with $G(\mathcal{O}^{A}) = 2a+b$ independent generators which commute
with each other except $\{ \bar{X}_{i} , \bar{Z}_{i} \} = 0$ ($i=1 ,
\cdots , a$). Therefore, independent generators are separated into the
commuting generators $\bar{X}_{a+1} \cdots \bar{X}_{a+b}$ and the
anti-commuting generators $\bar{X}_{1} , \cdots , \bar{X}_{a}$ and
$\bar{Z}_{1}, \cdots , \bar{Z}_{a}$. We call $G(\mathcal{O}^{A})$
independent generators which satisfy these commutation relations
\textit{canonical generators} for $\mathcal{O}^{A}$. Later, we shall
show that any subgroup of Pauli operator group $\mathcal{P}$ can be
represented through canonical generators. For simplicity, we introduce
a symbolic notation for canonical generators of $\mathcal{O}^{A}$
represented as
\begin{align}
\mathcal{O}^{A} =\left\langle
\begin{array}{cccccc}
   \bar{X}_{1} ,    &     \cdots , & \bar{X}_{a}     , & \bar{X}_{a+1}, & \cdots  ,& \bar{X}_{a+b}     \\
   \bar{Z}_{1}, &     \cdots , & \bar{Z}_{a} , & \phi     , & \cdots  ,& \phi  
\end{array}
 \right\rangle.
\end{align}
Here, generators in the same column anti-commute while any other pair
of generators in the different columns commute each other. The
\textit{null operator} $\phi$ below $\bar{X}_{i}$ means that
$\bar{X}_{i}$ has \textit{no anti-commuting operator pair} in
$\mathcal{O}^{A}$.

The commutation relations between canonical generators can be
concisely captured by considering them as if they act like Pauli
operators $X_{i}$ and $Z_{i}$ defined for each of single qubits. To
grasp commutation relations between canonical generators more clearly,
it is convenient to consider a Clifford transformation $U$ which
transforms $\mathcal{O}^{A}$ to a group of operators whose generators
are represented only with local Pauli operators $X_{i}$ and
$Z_{i}$. Let us consider the Clifford transformation $U$ obeying the
following conditions
\begin{align}
\bar{X}_{i} &= U X_{i} U^{\dagger} &(1\leq i\leq a+b)\\
\bar{Z}_{i} &= U Z_{i} U^{\dagger} &(1\leq i\leq a)    
\end{align}
and define a group $\mathcal{Q}$ as
\begin{align}
\mathcal{Q} =\left\langle
\begin{array}{cccccc}
  X_{1} ,    &     \cdots , & X_{a}     , & X_{a+1}, & \cdots  ,& X_{a+b}     \\
  Z_{1}, &     \cdots , & Z_{a} , & \phi     , & \cdots  ,& \phi  
\end{array}
 \right\rangle.
\end{align}
Then, $\mathcal{O}^{A}$ can be represented through the Clifford transformation $U$ as
\begin{align}
\mathcal{O}^{A} = U \mathcal{Q} U^{\dagger}
\end{align}
where the Clifford transformation $U$ acts on each of elements inside the groups of operators.

\subsection{Derivation of Canonical Generators}
\label{sec:3D}

Now, let us provide an explicit procedure to obtain canonical
generators, not only for the overlapping operator group, but also for
an arbitrary subgroups of the Pauli operator group $\mathcal{O} \in
\mathcal{P}$. Before describing each step to obtain canonical
generators in detail, let us begin by looking at some simple examples.

As the most familiar example, the centralizer group $\mathcal{C}$ is
represented in a canonical form as
\begin{align}
\mathcal{C} =\left\langle
\begin{array}{cccccc}
 \ell_{1} , & \cdots , &  \ell_{k} , & S_{1} ,    &     \cdots , & S_{G(\mathcal{S})}         \\
 r_{1} , & \cdots , &  r_{k} , & \phi , & \cdots  ,& \phi  
\end{array}
 \right\rangle 
\end{align}
where $S_{1}, \cdots , S_{G(\mathcal{S})}$ represent $G(\mathcal{S})$
independent stabilizer generators. One can easily see that logical
operators commute with all the stabilizer generators, but not inside
the stabilizer group $\mathcal{S}$ in this representation.

Next, for $\mathcal{O} = \langle O_{1}, O_{2}, iI \rangle$ generated
from two commuting Pauli operators $O_{1}$ and $O_{2}$ with
$[O_{1},O_{2}]=0$, we can set $\bar{X}_{1} = O_{1} $ and $\bar{X}_{2}
= O_{2} $. Then, $\mathcal{O}$ is represented as
\begin{align}
\mathcal{O} =\left\langle
\begin{array}{cc}
  O_{1}  ,& O_{2}   \\
  \phi   ,& \phi
\end{array}
 \right\rangle.
\end{align}
On the other hand, if $O_{1}$ and $O_{2}$ are anti-commuting, we can
set $\bar{X}_{1} = O_{1} $ and $\bar{Z}_{1} = O_{2} $ and
$\mathcal{O}$ is represented as
\begin{align}
\mathcal{O} =\left\langle
\begin{array}{c}
  O_{1}   \\
  O_{2}  
\end{array}
 \right\rangle.
\end{align}

Now let us consider a more complicated example, where $\mathcal{O} =
\langle O_{1}, O_{2}, O_{3}, iI \rangle$ is generated from three
independent generators with commutations relations characterized as
\begin{equation}
\begin{split}
[O_{1},O_{2}]&=0 \\
\{O_{1},O_{3}\} &=0 \\
\{O_{2} , O_{3}\} &=0.
\end{split}
\end{equation}
We start by representing a group of operators $\mathcal{O}' = \langle
O_{1}, O_{2}, iI \rangle$ as
\begin{align}
\mathcal{O}' =\left\langle
\begin{array}{cc}
  O_{1} , & O_{2}  \\
  \phi  , & \phi  
\end{array}
 \right\rangle.
\end{align}
Though we might be tempted to put $O_{3}$ below $O_{1}$ in place of
the null operator $\phi$ due to anti-commutation between $O_{1}$ and
$O_{3}$, we soon notice that $O_{3}$ also anti-commutes with
$O_{2}$. Therefore, by replacing $O_{2}$ with $O_{1}O_{2}$, we obtain
the canonical representation of $\mathcal{O}$ as
\begin{align}
\mathcal{O} =\left\langle
\begin{array}{cc}
  O_{1} , & O_{1}O_{2}  \\
  O_{3}  ,& \phi  
\end{array}
 \right\rangle.
\end{align}

Finally, here is a general procedure to obtain canonical generators
for an arbitrary group of operators $\mathcal{O}= \langle iI , \{O_{i}
, \hspace{1ex} \forall i\} \rangle$. Let us suppose that an operator
group $\mathcal{M}$ is already represented in a canonical form as
\begin{align}
\mathcal{M} =\left\langle
\begin{array}{cccccc}
   \bar{X}_{1} ,    &     \cdots , & \bar{X}_{a}     , & \bar{X}_{a+1}, & \cdots  ,& \bar{X}_{a+b}     \\
   \bar{Z}_{1}, &     \cdots , & \bar{Z}_{a} , & \phi     , & \cdots  ,& \phi  
\end{array}
 \right\rangle.
\end{align}
We derive a canonical representation for an operator group
$\mathcal{M}' = \langle \mathcal{M}, U, iI \rangle \supset
\mathcal{M}$ with $U \not \in \mathcal{M}$. Therefore, by iterating
the procedure we shall describe in the below, one can represent any
subgroup of the Pauli operator group in canonical representations. Let
us represent the commutation relations between $U$ and the original
canonical generators as
\begin{align}
\bar{X}_{i} U &= (-1)^{p_{i}} U \bar{X}_{i}   &(1\leq i\leq a)   \\
\bar{Z}_{i} U &= (-1)^{q_{i}} U \bar{Z}_{i}   &(1\leq i\leq a)      \\
\bar{X}_{a+i} U &= (-1)^{t_{i}} U \bar{X}_{a+i}   &(1\leq i\leq b)
\end{align}
with $p_{i},q_{i},t_{i}=0 ,1$ where $0$ represents the commutations
between $U$ and corresponding canonical generators while $1$
represents the anti-commutations. Here, we define a new operator
\begin{align}
U'= \prod_{i} \bar{X}_{i}^{q_{i}} \bar{Z}_{i}^{p_{i}} U
\end{align}
such that $U'$ commutes with $\bar{X}_{i}$ and $\bar{Z}_{i}$ for
$\forall i$. This can be easily verified by directly checking the
commutation relations
\begin{equation}
\begin{split}
[U', \bar{X}_{i}]&= [ \bar{Z}_{i}^{p_{i}} U   ,\bar{X}_{i} ] = 0 \\
[U', \bar{Z}_{i}]&= [ \bar{X}_{i}^{q_{i}} U   ,\bar{Z}_{i} ] = 0 .
\end{split}
\end{equation}
First, if $t_{i}=0$ for $\forall i$, we have the canonical representation of $\mathcal{M}'$ as
\begin{align}
\mathcal{M}' =\left\langle
\begin{array}{ccccccc}
   \bar{X}_{1} ,    &     \cdots , & \bar{X}_{a}     , & \bar{X}_{a+1}, & \cdots  ,& \bar{X}_{a+b} , & U'   \\
   \bar{Z}_{1}, &     \cdots , & \bar{Z}_{a} , & \phi     , & \cdots  ,& \phi  , & \phi
\end{array}
 \right\rangle
\end{align}
since $U'$ commutes with all the operators in $\mathcal{M}$. Next, we
consider the case where there exist integers $j$ with $t_{j}=1$ ($j
\leq b$). Without loss of generality, we can assume that
$t_{1}=1$. Then, the canonical representation of $\mathcal{M}$ is
\begin{align}
\mathcal{M}' =\left\langle
\begin{array}{cccccc}
   \bar{X}_{1} ,    &     \cdots , & \bar{X}_{a}     , & \bar{X}_{a+1}' , & \bar{X}_{a+2}' , & \cdots  \\
   \bar{Z}_{1},     &     \cdots , & \bar{Z}_{a}     , & U'     , & \phi  ,& \cdots  
\end{array}
 \right\rangle
\end{align}
where 
\begin{align}
\bar{X}_{a+i}' &= \bar{X}_{a+i} \bar{X}_{a+1}  &(  r_{i}=1, \hspace{1ex} i \not = 1 ) \\
\bar{X}_{a+i}' &= \bar{X}_{a+i} &(r_{i} = 0).
\end{align} 
We can easily see that $[\bar{X}_{i}', U]=0$ for $\forall i$ by
directly checking the commutation relations. Therefore, given the
canonical representation of $\mathcal{M}^{'}$, we can find the
canonical representation of $\mathcal{M}$ by following the above
procedure. By treating each $O_{i}$ as $U$, we can obtain canonical
generators for $\mathcal{O}= \langle iI , \{O_{i} , \hspace{1ex}
\forall i\} \rangle$.

Note that this procedure can be viewed as a generalization of Gaussian
elimination, applied to commuting subgroups of the Pauli operator
group \cite{Stab2}, but now made effective also for a group of
operators consisting of Pauli operators which may not commute each other. 
%
%
The above procedure clearly defines a general procedure to obtain
canonical generators for any subgroup of the Pauli operator group
$\mathcal{P}$. Moreover, in contrast to the computational cost of the
``exhaustive elimination'' approach, which scales with $|S|$, the
computational cost of this procedure grows as $\sim\log(|S|)$.
Specifically, each of the iteration steps can be performed in a number
of steps which is of the order of $G(\mathcal{M})$. Therefore, in
total, this procedure only requires a number of steps which scales
linearly in $G(\mathcal{O})^{2}$.

We emphasize that the choices of canonical generators are not unique
for a given group of operator $\mathcal{O}$. First, note that the
largest Abelian subgroup of $\mathcal{O}$ is uniquely defined as
\begin{align}
\mathcal{O}_{c} = \langle \bar{X}_{a+1},  \cdots  , \bar{X}_{a+b} \rangle.
\end{align}
Therefore, the commuting canonical generators $\bar{X}_{a+1}, \cdots ,
\bar{X}_{a+b}$ can be chosen freely from $\mathcal{O}_{c}$ as long as
each of them is independent. Also, there is a freedom to apply an
element from $\mathcal{O}_{c}$ to the anti-commuting canonical
generators $\bar{X}_{1}, \cdots, \bar{X}_{a}$ and $\bar{Z}_{1},
\cdots, \bar{Z}_{a}$ since this does not change the commutation
relations. Finally, for two given pairs of anti-commuting canonical
generators $\bar{X}_{1}$, $\bar{Z}_{1}$, $\bar{X}_{2}$ and
$\bar{Z}_{2}$, the following choices of anti-commuting canonical
generators
\begin{equation}
\begin{split}
\bar{X}_{1} &\leftrightarrow \bar{X}_{1}\bar{X}_{2} \\
\bar{Z}_{1} &\leftrightarrow \bar{Z}_{1} \\
\bar{X}_{2} &\leftrightarrow \bar{X}_{2} \\
\bar{Z}_{2} &\leftrightarrow \bar{Z}_{1}\bar{Z}_{2} 
\end{split}
\end{equation}
also satisfy the commutation relations of canonical generators. This
arbitrariness in the choices of anti-commuting canonical generators
leads to the arbitrariness of the definitions of logical qubits. We
will treat this problem carefully when non-local properties of logical
qubits are discussed in section~\ref{sec:discussion}.

\section{Framework \romannumeral 2}
\label{sec:logical}

In the previous section, we described the classification of logical
operators based on their localities and non-localities. Then, we
defined the overlapping operator group and presented a theoretical
tool, the canonical representation, to analyze the overlapping
operator group. In this section, we connect the classification of
logical operators with the overlapping operator group by using the
canonical representation to compute logical operators in each of four
sets $M_{A}$, $M_{B}$, $M_{AB}$ and $M_{\phi}$.

Computation of logical operators in $M_{A}$, $M_{B}$ and $M_{AB}$ can
be performed by directly analyzing the overlapping operator groups
$\mathcal{O}^{A}$ and $\mathcal{O}^{B}$ through a canonical
representation, as we see in section \ref{sec:4A} and
\ref{sec:4B}. For the computation of logical operators in $M_{\phi}$,
the relation between $\mathcal{O}^{A}$ and $\mathcal{O}^{B}$ needs to
be analyzed. We show that there is a one-to-one correspondence between
anti-commuting canonical generators for $\mathcal{O}^{A}$ and
$\mathcal{O}^{B}$, in section \ref{sec:4C}. This correspondence leads
to an efficient means to compute the logical operators in $M_{\phi}$,
shown in section \ref{sec:4D}.

\subsection{Localized Logical Operators; $M_{AB}$}
\label{sec:4A}

Let us begin by obtaining all the localized logical operators in
$M_{AB}$, through the canonical representation of the overlapping
operator group $\mathcal{O}^{A}$. We can represent the overlapping
operator group $\mathcal{O}^{A}$ as
\begin{align}
\mathcal{O}^{A} =\left\langle
\begin{array}{cccccc}
   \bar{X}_{1} ,    &     \cdots , & \bar{X}_{a}     , & \bar{X}_{a+1}, & \cdots  ,& \bar{X}_{a+b}     \\
   \bar{Z}_{1}, &     \cdots , & \bar{Z}_{a} , & \phi     , & \cdots  ,& \phi  
\end{array}
 \right\rangle.
\end{align}
Since $\mathcal{S}_{A}\subseteq \mathcal{O}^{A}$ and can be represented as
\begin{align}
\mathcal{S}_{A} = \langle S_{1} ,  \cdots , S_{G(\mathcal{S}_{A})} \rangle  
\end{align}
with $G(\mathcal{S}_{A})$ independent stabilizer generators, we can represent $\mathcal{O}^{A}$ as
\begin{align}
\mathcal{O}^{A} &=
\left\langle
\begin{array}{ccccccccc}
 \bar{X}_{1}  ,   &     \cdots , & \bar{X}_{a}     ,S_{1} , & \cdots , & S_{G(\mathcal{S}_{A})}, & \ell_{1}, & \cdots ,& \ell_{d}    \\
 \bar{Z}_{1}, &     \cdots , & \bar{Z}_{a} ,\phi  , & \cdots , & \phi ,                  & \phi, & \cdots , & \phi      
\end{array}
 \right\rangle
\end{align}
by extracting stabilizer generators for $\mathcal{S}_{A}$. We may then
note that operators $\ell_{1}, \cdots , \ell_{d}$ commute with all the
operators in $\mathcal{O}^{A}$, but not inside
$\mathcal{S}_{A}$. Therefore, operators $\ell_{1}, \cdots , \ell_{d}$ are
actually \textit{localized logical operators in $L_{A}$}.

\begin{figure}[ht!]
\begin{center}
\includegraphics[width=0.9\linewidth]{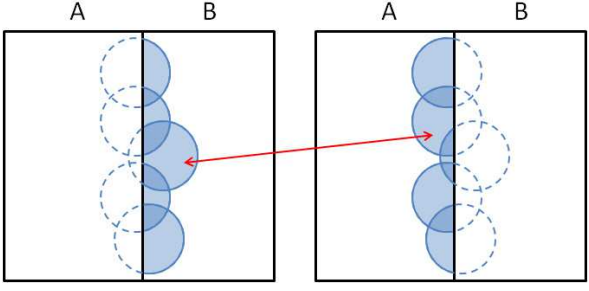}
\caption{Two equivalent logical operators in $M_{AB}$ found both
  inside $\mathcal{O}^{A}$ and $\mathcal{O}^{B}$.  Each circle
  represents stabilizer generators which overlap at the boundary
  between $A$ and $B$.}
\label{counterpart}
\end{center}
\end{figure}

In fact, we can easily see that logical operators $\ell_{1}, \cdots ,
\ell_{d}$ are in $M_{AB}$ by representing $\ell_{1}, \cdots , \ell_{d}$ as
\begin{align}
\ell_{i} = \prod_{j \in R_{i}} S_{j}|_{A} \in \mathcal{O}^{A}
\end{align}
for some sets of integers $R_{i}$ for $i=1 , \cdots ,d$. Then, there
exist localized logical operators
\begin{align}
\ell_{i}' = \prod_{j \in R_{i}} S_{j}|_{A} S_{j} = \prod_{j \in R_{i}} S_{j}|_{B} \in \mathcal{O}^{B}
\end{align}
which are now defined inside $B$, but equivalent to $\ell_{i}$
(Fig.\ref{counterpart}). Therefore, $\ell_{1}, \cdots , \ell_{d}$ are in
$M_{AB}$. Soon, we shall see that logical operators $\ell_{1}, \cdots ,
\ell_{d}$ are all the independent logical operators in $M_{AB}$.

\subsection{Localized Logical Operators; $M_{A}$ and $M_{B}$}
\label{sec:4B}

Though we obtained localized logical operators in $M_{AB}$ through the
analysis of the overlapping operator groups, this approach does not
exhaust all the localized logical operators in $L_{A}$. In order to
find logical operators in $M_{A}$, let us represent the restriction of
the centralizer group into $A$ in a canonical form as
\begin{align}
\mathcal{C}_{A} &= \langle \{ U \in \mathcal{P}_{A} \mid [U, O]=0,\hspace{1ex} O \in \mathcal{O}^{A} \} \rangle \\ \notag
 &=
\left\langle
\begin{array}{ccccccccc}
S_{1} , & \cdots , & S_{G(\mathcal{S}_{A})}, & \ell_{1}, & \cdots ,& \ell_{d} ,& \alpha_{1}, & \cdots ,& \alpha_{c}   \\
\phi  , & \cdots   , & \phi                  , & \phi ,&  \cdots , & \phi , & \alpha_{1}' , & \cdots , & \alpha_{c}'         
\end{array}
 \right\rangle.
 \end{align}
Thus, we immediately notice that $\alpha_{1} \cdots \alpha_{c}$ and
$\alpha_{1}' \cdots \alpha_{c}'$ are localized logical operators in
$M_{A}$. Since $\mathcal{C}_{A}$ includes all the operators defined
inside $A$ which commute with all the stabilizer generators, $\ell_{1},
\cdots , \ell_{d}$, $\alpha_{1} \cdots \alpha_{c}$ and $\alpha_{1}'
\cdots \alpha_{c}'$ are all the localized logical operators in
$L_{A}$.

Now we show that $\ell_{1}, \cdots , \ell_{d}$ are all the independent
logical operators in $M_{AB}$ while $\alpha_{1} \cdots \alpha_{c}$ and
$\alpha_{1}' \cdots \alpha_{c}'$ are logical operators in
$M_{A}$. Since logical operators in $M_{AB}$ can be defined both
inside $A$ and $B$, they must commute with all the localized logical
operators in $L_{A}$. Therefore, logical operators from $\alpha_{1}
\cdots \alpha_{c}$ and $\alpha_{1}' \cdots \alpha_{c}'$ cannot be
inside $M_{AB}$ since they have anti-commuting pairs inside
$L_{A}$. Thus, we notice that $\ell_{1}, \cdots , \ell_{d}$ are in $M_{AB}$,
while $\alpha_{1} \cdots \alpha_{c}$ and $\alpha_{1}' \cdots
\alpha_{c}'$ are in $M_{A}$.

\subsection{One-to-one Correspondence between $\mathcal{O}^{A}$ and $\mathcal{O}^{B}$}
\label{sec:4C}

Unlike for the localized logical operators obtained above, the
computation of non-local logical operators is somewhat complicated,
since we need to discuss the commutation relations of operators inside
both $A$ and $B$ at the same time. Therefore, we need a more
sophisticated approach, which allows us to discuss two overlapping
operator groups $\mathcal{O}_{A}$ and $\mathcal{O}_{B}$
simultaneously.

There is an intrinsic one-to-one correspondence between anti-commuting
canonical generators in $\mathcal{O}_{A}$ and
$\mathcal{O}_{B}$. Surprisingly, this correspondence between
$\mathcal{O}_{A}$ and $\mathcal{O}_{B}$ enables us to compute all the
non-local logical operators \textit{only through the computations
  performed locally inside each of complementary subsets $A$ and $B$}.

First, let us recall the canonical representation for the overlapping
operator group $\mathcal{O}^{A}$ with
\begin{align}
\mathcal{O}^{A} &=
\left\langle
\begin{array}{ccccccc}
 \bar{X}_{1}  ,   &     \cdots , & \bar{X}_{a}     ,& \ell_{1}, & \cdots ,& \ell_{d} ,& \mathcal{S}_{A}   \\
 \bar{Z}_{1}, &     \cdots , & \bar{Z}_{a} ,& \phi  , & \cdots , & \phi ,                  & \phi     
\end{array}
 \right\rangle
\end{align}
where $\mathcal{S}_{A}$ represents the generators for
$\mathcal{S}_{A}$ symbolically. We can represent $\bar{X}_{i}$,
$\bar{Z}_{i}$ and $\ell_{i}$ as
\begin{align}
\bar{X}_{i} &= \prod_{j \in R^{(x)}_{i}} S_{j}|_{A} \in \mathcal{O}^{A} \\
\bar{Z}_{i} &= \prod_{j \in R^{(z)}_{i}} S_{j}|_{A} \in \mathcal{O}^{A} \\
\ell_{i} &= \prod_{j \in R_{i}} S_{j}|_{A} \in \mathcal{O}^{A}.
\end{align}
through the overlaps $S_{i}|_{A}$ and some sets of integers
$R^{(x)}_{i}$, $R^{(x)}_{i}$ and $R_{i}$. Then, we notice that the
operators
\begin{align}
\bar{X}_{i}' &= \prod_{j \in R^{(x)}_{i}} S_{j}|_{B} \in \mathcal{O}^{B} \\
\bar{Z}_{i}' &= \prod_{j \in R^{(z)}_{i}} S_{j}|_{B} \in \mathcal{O}^{B} \\
\ell_{i}'       &= \prod_{j \in R_{i}} S_{j}|_{B} \in \mathcal{O}^{B}
\end{align}
actually form canonical generators for the overlapping operator group
$\mathcal{O}^{B}$. In fact, simple calculations lead to that
\begin{align}
\mathcal{O}^{B} &=
\left\langle
\begin{array}{ccccccc}
 \bar{X}_{1}'  ,   &     \cdots , & \bar{X}_{a}'     ,& \ell_{1}', & \cdots ,& \ell_{d}' ,& \mathcal{S}_{B}   \\
 \bar{Z}_{1}', &     \cdots , & \bar{Z}_{a}' ,& \phi  , & \cdots , & \phi ,                  & \phi     
\end{array}
 \right\rangle.
\end{align}
We can easily see that $\bar{X}_{i}'$, $\bar{Z}_{i}'$ and $\ell_{i}'$ are
independent generators since we can also construct the canonical
generators for $\mathcal{O}^{A}$ by starting from the canonical
generators for $\mathcal{O}^{B}$. This construction also ensures that
there are the same numbers of anti-commuting canonical generators for
$\mathcal{O}^{A}$ and $\mathcal{O}^{B}$. We can easily check the
commutation relations between $\bar{X}_{i}'$, $\bar{Z}_{i}'$ and
$\ell_{i}'$ which are exactly the same as the commutation relations
between $\bar{X}_{i}$, $\bar{Z}_{i}$ and $\ell_{i}$. Therefore, there is
a \textit{one-to-one correspondence} between canonical generators for
$\mathcal{O}^{A}$ and $\mathcal{O}^{B}$.

We can see the origin of this correspondence by extracting
$\mathcal{S}_{A}$ and $\mathcal{S}_{B}$ from the stabilizer group
$\mathcal{S}$ as
\begin{align}
\mathcal{S}= \langle \mathcal{S}_{A} , \mathcal{S}_{B} , \mathcal{S}_{AB} \rangle
\end{align}
where the \textit{non-local stabilizer group} $\mathcal{S}_{AB}$ satisfies
\begin{align}
G(\mathcal{S}) = G(\mathcal{S}_{A}) + G(\mathcal{S}_{B}) + G(\mathcal{S}_{AB}).
\end{align}
All the generators for $\mathcal{S}_{AB}$ are non-locally defined
jointly over $A$ and $B$. Here, we note that the definition of the
non-local stabilizer group $\mathcal{S}_{AB}$ is not unique since we
have a freedom to apply stabilizers inside $\mathcal{S}_{A}$ and
$\mathcal{S}_{B}$ to generators for $\mathcal{S}_{AB}$. Now, we notice
that the following stabilizer generators are all the independent
generators for $\mathcal{S}_{AB}$;
\begin{align}
S^{(x)}_{i} &\equiv \prod_{j \in R^{(x)}_{i}} S_{j} =
\bar{X}_{i}\bar{X}_{i}' \\ S^{(z)}_{i} &\equiv \prod_{j \in
  R^{(z)}_{i}} S_{j} = \bar{Z}_{i}\bar{Z}_{i}' \\ S^{(L)}_{i} &\equiv
\prod_{j \in R_{i}} S_{j} = \ell_{i}\ell_{i}'.
\end{align}

Note that the analysis of the overlapping operator group
$\mathcal{O}^{A}$ automatically generates anti-commuting canonical
generators for the overlapping operator group $\mathcal{O}^{B}$. In
particular, a one-to-one correspondence we revealed here provides
independent generators for $\mathcal{S}_{AB}$ explicitly.

\subsection{Non-Local Logical Operators; $M_{\phi}$}
\label{sec:4D}

Having uncovered the relations between $\mathcal{O}^{A}$ and
$\mathcal{O}^{B}$, let us finally obtain non-local logical operators.
First, we define a group of operators which commute with all the
stabilizer generators inside $A$ as
\begin{align}
\mathcal{C}(\mathcal{S}_{A}) = \langle \{ U \in \mathcal{P}_{A} \mid [U, O]=0 ,\hspace{1ex} \forall O \in \mathcal{S}_{A}  \}     \rangle 
\end{align}
and a similar group of operators inside $B$ as  
\begin{align}
\mathcal{C}(\mathcal{S}_{B}) = \langle \{ U \in \mathcal{P}_{B} \mid [U, O]=0 ,\hspace{1ex} \forall O \in \mathcal{S}_{B}  \}     \rangle .
\end{align}
Their canonical representations can be obtained as
\begin{align}
\mathcal{C}(\mathcal{S}_{A}) &=
\left\langle
\begin{array}{cccccc}
 \{ \bar{X}_{i} \} &, \mathcal{S}_{A} , & \ell_{1}, & \cdots ,& \ell_{d} ,& \{ \alpha_{i} \}   \\
 \{ \bar{Z}_{i} \} &, \phi            , & r_{1}, & \cdots ,& r_{d} ,& \{ \alpha_{i}' \}        
\end{array}
 \right\rangle
\end{align}
and
\begin{align}
\mathcal{C}(\mathcal{S}_{B}) &=
\left\langle
\begin{array}{cccccc}
 \{ \bar{X}_{i}' \} &, \mathcal{S}_{B} , & \ell_{1}', & \cdots ,& \ell_{d}' ,& \{ \beta_{i} \}   \\
 \{ \bar{Z}_{i}' \} &, \phi            , & r_{1}', & \cdots ,& r_{d}' ,& \{ \beta_{i}' \}        
\end{array}
 \right\rangle
\end{align}
where $r_{i}$ and $r_{i}'$ are anti-commuting pairs of $\ell_{i}$ and
$\ell_{i}'$ defined inside $A$ and $B$ respectively. Here, we used $\{
\bar{X}_{i} \}$ symbolically to represent canonical generators
$\bar{X}_{1}, \cdots, \bar{X}_{a}$.

Now, let us consider the following operators $\delta_{i}$ 
\begin{align}
\delta_{i} &= r_{i}r_{i}'
\end{align}
which are defined jointly over $A$ and $B$ for $i=1,\cdots, d$. Let us
define the set of $d$ different $\delta_{i}$ as
\begin{align}
\Delta = \{ \delta_{1}, \cdots , \delta_{d} \} \label{delta}.
\end{align}
We may now show that the set of jointly defined operators $\Delta$
consists only of non-local logical operators in $M_{\phi}$. First, we
can easily see that $\delta_{i}$ commute with all the stabilizers in
$\mathcal{S}_{A}$ and $\mathcal{S}_{B}$. We can also check the
commutations with stabilizer generators in $\mathcal{S}_{AB}$ by
seeing
\begin{align}
[\delta_{i}, S^{(x)}_{i}] &= [r_{i}r_{i}', \bar{X}_{i}\bar{X}_{i}']=0 \\
[\delta_{i}', S^{(z)}_{i}] &= [r_{i}r_{i}', \bar{Z}_{i}\bar{Z}_{i}']=0 \\
[\delta_{i}', S^{(L)}_{i}] &= [r_{i}r_{i}', \ell_{i}\ell_{i}']=0.
\end{align}
Since $\delta_{i}$ commute with all the stabilizer generators, but not
inside the stabilizer group $\mathcal{S}$, $\delta_{i}$ are logical
operators. We can show that $\delta_{i}$ are in $M_{\phi}$ by
considering the anti-commutations with $\ell_{i}$
\begin{align}
\{\delta_{i}, \ell_{i}\} = \{\delta_{i}, \ell_{i}' \} = 0.
\end{align}
Suppose that $\delta_{i}$ can be also defined inside $A$. Then,
$\delta_{i}$ commute with all the logical operators in $M_{AB}$ which
leads to a contradiction. Therefore, $\delta_{i}$ in $\Delta$ are
non-local logical operators in $M_{\phi}$. Later, we shall see that
these $\delta_{i}$ are all the independent non-local logical
operators. Since the proof requires further analysis on localized
logical operators, we postpone it until section \ref{geometry}.

Here, let us mention the reduction of computational cost of obtaining non-local logical operators as a result of the one-to-one correspondence between canonical generators between the overlapping operator groups $\mathcal{O}^{A}$ and $\mathcal{O}^{B}$. For the computations of localized logical operators, one only needs to see whether a given operator commutes with the overlaps of the stabilizer generators, which can be efficiently checked through our analysis based on the overlapping operator group. However, for the computations of non-local logical operators, one needs to discuss the commutation relationships with stabilizer generators defied globally over a bi-partition since non-local logical operators cannot be defined locally. Though a naive approach is to check the commutation relations for all the stabilizer generators, due to the one-to-one correspondence between the overlapping operator groups $\mathcal{O}^{A}$ and $\mathcal{O}^{B}$, we only need to consider the commutation relations inside each of subsets $A$ and $B$.

\begin{figure}[ht!]
\begin{center}
\includegraphics[width=0.8\linewidth]{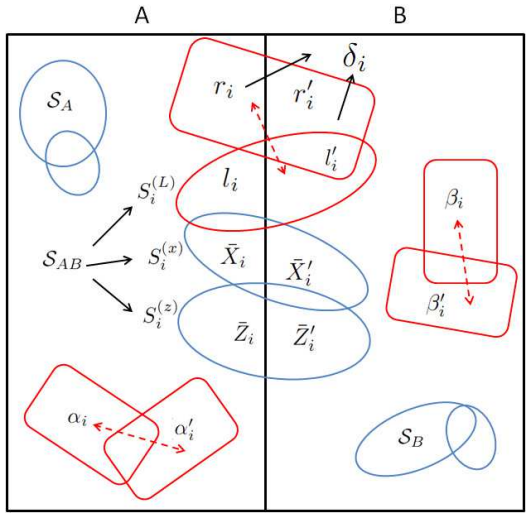}
\caption{An illustration of stabilizer generators and logical
  operators in a bi-partition. Circles represent stabilizer generators
  while squares represent logical operators except that $\ell_{i}$ and
  $\ell_{i}'$ are found inside stabilizer generators.  The dotted arrows
  show the possibility of anti-commutations.  Logical operators
  $\alpha_{i}$ and $\alpha_{i}'$ ($\beta_{i}$ and $\beta_{i}'$) in
  $M_{A}$ ($M_{B}$) form pairs just inside $A$ ($B$).  Stabilizer
  generators for non-local stabilizer group $\mathcal{S}_{AB}$ are
  represented as products of canonical generators for
  $\mathcal{O}^{A}$ and $\mathcal{O}^{B}$.  While logical operators
  $\ell_{i}$ and $\ell_{i}'$ in $M_{AB}$ appear as the projection of some
  stabilizer generators $S_{i}^{(L)}$ for $\mathcal{S}_{AB}$,
  non-local logical operators $\delta_{i}$ are products of
  anti-commuting pairs of $\ell_{i}$ and $\ell_{i}'$.}
\label{summary}
\end{center}
\end{figure}

We summarize our results graphically in
Fig.\ref{summary} which shows all the logical operators along with the
canonical generators for the overlapping operator groups. The
procedures to compute logical operators in $M_{A}$, $M_{B}$, $M_{AB}$
and $M_{\phi}$ described in this section complete our framework.

\section{Examples: Non-Local Properties of Stabilizer Codes}
\label{sec:discussion}

The two sections above present a framework providing computational
tractable means for computing and classifying logical operators based
on their non-localities and localities. In particular, this framework
includes procedures to derive all the logical operators in each of
four sets $M_{AB}$, $M_{A}$, $M_{B}$ and $M_{\phi}$ through the
analysis of the overlapping operator groups, using canonical
representations.  We now turn to a demonstration of the usefulness of
this framework, by studying some interesting non-local properties of
the stabilizer formalism.  Below, we consider global symmetries in
topological order, the distribution of multi-partite entanglement, and
entanglement entropy.

\subsection{Geometric duality and topological order \label{geometry}}

A valuable application of our framework is the study of geometric
properties of logical operators. The framework has required no
geometry so far, since our discussion has concentrated on properties
of logical operators and stabilizers in the Pauli operator space,
without considering the geometric locations of qubits in real physical
space. Though the framework discussed locality and non-locality of
logical operators, the bi-partition was placed inside the Pauli
operator group $\mathcal{P}$. However, the notion of geometry becomes
particularly important, for example, when the practical implementation
of quantum codes is the issue. When quantum codes are defined on some
geometric manifold, a bi-partition in the Pauli operator space becomes
a bi-partition in real physical space consisting of physical qubits.

An interesting problem where geometries of logical operators become
essential is the study of topological order emerging in quantum
codes. The mathematical notion of {\em topology} is widely appreciated
as playing a crucial role in many interesting physical phenomena
\cite{Z2_insulator, Moore91, Read00, Nayak08, Wen_Text}, including
some quantum codes \cite{Toric}. Topological order also possesses
potential practical importance, since a system with topological order
may support dissipationless currents \cite{Murakami03} and serve as a
resource for universal quantum computation
\cite{Quantum_Double}. Topological properties of various systems have
been analyzed using a range of quantities, including Berry phases
\cite{Shapere_Wilczek}, anyonic excitations, and topological entropy
\cite{Topo1, Topo2}.

Topological order is commonly believed to be a result of global
symmetries in the system \cite{Wen_Text}. Such a global symmetry
emerges as the existence of global operators which commute with the
system Hamiltonian. If this observation is interpreted in the language
of quantum codes, it asserts that the existence of large logical
operators implies the existence of topological order in the
system. However, why large logical operators may give rise to
topological order is not well understood. To answer this fundamental
question about topological order and symmetries, topological aspects
of logical operators need to be analyzed. Thus, the analysis of the
geometric invariance (and variance) of shapes of logical operators is
a necessary step to understand an intrinsic connection between global
symmetries and topological order.

Here, we add geometry to the discussion of quantum codes by proving a
theorem which indicates the existence of an intrinsic duality on the
geometric shapes of logical operators in stabilizer codes. This
theorem can be directly proven using our framework. We apply this
theorem to the Toric code \cite{Toric} which supports topological
order in the ground state space. We give a general discussion on the
Toric code from a viewpoint of symmetries of the system by analyzing
geometric properties of logical operators.

The theorem we prove in this subsection also complements our
framework. Using this theorem, we show that non-local logical
operators $\delta_{i}$ in $\Delta$ in section~\ref{sec:4D} are all the
possible independent non-local logical operators.

\subsubsection{Duality of logical operators in a bi-partition}

Our main goal in this subsection is to discuss the geometry of logical
operators. Consider a stabilizer code defined with some physical
qubits on some geometric manifold. The geometric shape of a logical
operator may be simply captured by considering the shape of qubits
where the logical operator has non-trivial support. However, logical
operators have many equivalent representations, since the application
of stabilizers does not change the properties of logical
operators. This makes it difficult to uniquely define the geometry of
a logical operator, or to determine the geometrically invariant
properties of a given logical operator.

To avoid this difficulty, we take the following approach. If a logical
operator is defined inside some localized region $A$, then, the shape
of the logical operator may be bounded by the shape of $A$. By
analyzing the geometric shapes of possible regions where each logical
operator can be supported, geometric properties of logical operators
may be studied. For this purpose, localized logical operators obtained
in our framework become essential. We start our discussion of
geometries of logical operators by proving the following theorem
governing localized logical operators in a bi-partition. This theorem
can be also be easily derived from the discussions independently done
elsewhere \cite{Wilde09}.

\begin{theorem}\label{theorem1}
For a stabilizer code with $k$ logical qubits, let $g_{A}$ and $g_{B}$
be the numbers of independent logical operators which can be supported
only by qubits inside $A$ and $B$ respectively with $B= \bar{A}$. Then
the following formula holds:
\begin{align}
g_{A} + g_{B} = 2k.
\end{align}
\end{theorem}

Theorem~\ref{theorem1} states that by naively counting the number of
independent logical operators which can be defined inside each region
and taking the sum of them, the sum is equal to $2k$. Therefore, the
sum of the number of localized logical operators inside each subset
$A$ and $B$ is always conserved. The proof is immediate with our
framework.

\begin{figure}[ht!]
\begin{center}
\includegraphics[width=0.9\linewidth]{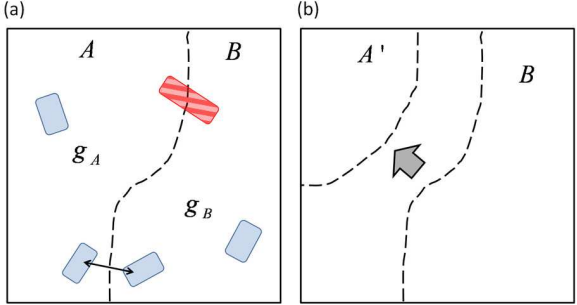}
\caption{(a) Geometric duality on stabilizer codes ($g_{A}+g_{B}=2k$).
  Rectangles represent logical operators. $g_{A}$ and $g_{B}$
  represent the number of localized logical operators inside $A$ and
  $B$. A hatched rectangle (red online) at the boundary between $A$
  and $B$ is a non-local logical operator which is not counted either
  in $g_{A}$ or in $g_{B}$. Rectangles connected by a two-sided arrow
  are equivalent and counted both in $g_{A}$ and in $g_{B}$. Other
  rectangles are localized logical operators defined only inside $A$
  or $B$ and counted once in $g_{A}$ or in $g_{B}$.  (b) Shrinking
  logical operators. When $g_{A} = g_{A'}$ while $A'$ is smaller than
  $A$, all the logical operators in $A$ have equivalent logical
  operators in $A'$. Application of appropriate stabilizers shrinks
  the shapes of logical operators from $A$ to $A'$.}
\label{bipartition_theorem}
\end{center}
\end{figure}

\begin{proof}
Since all the localized logical operators are defined inside
$\mathcal{C}_{A}$ and $\mathcal{C}_{B}$, we have
\begin{align}
g_{A} &= G(\mathcal{C}_{A}) - G(\mathcal{S}_{A})\\
g_{B} &= G(\mathcal{C}_{B}) - G(\mathcal{S}_{B}).
\end{align}
By looking at the relation between $\mathcal{O}^{A}$ and $\mathcal{C}_{A}$, we have
\begin{align}
G(\mathcal{C}_{A}) = 2V_{A} - G(\mathcal{O}_{A})
\end{align}
where $V_{A}$ is the number of qubits inside $A$.
Therefore, we have
\begin{equation}
\begin{split}
g_{A} + g_{B} = &2(V_{A} + V_{B}) - G(\mathcal{O}_{A}) - G(\mathcal{O}_{B})\\ 
                &- G(\mathcal{S}_{A}) -G(\mathcal{S}_{B}). 
\end{split}
\end{equation}
Now, by using the generators for $\mathcal{S}_{AB}$ represented
through canonical generators in $\mathcal{O}_{A}$ and
$\mathcal{O}_{B}$, we have
\begin{align}
G(\mathcal{O}^{A}) &= G(\mathcal{S}_{A}) + G(\mathcal{S}_{AB}) \\
G(\mathcal{O}^{B}) &= G(\mathcal{S}_{B}) + G(\mathcal{S}_{AB}).
\end{align}
Finally, from $V_{A} + V_{B}=N$ and $G(\mathcal{S}_{A}) +
G(\mathcal{S}_{B}) + G(\mathcal{S}_{AB}) = N-k$, we have
\begin{align}
g_{A} + g_{B} = 2k.
\end{align}
\end{proof}

Note that Theorem \ref{theorem1} is generally true for all the
stabilizer codes with any bi-partitions. Before introducing
geometries, we make a few remarks on Theorem \ref{theorem1}. First,
let us discuss the relationship of this theorem with our
classification of logical operators. $g_{A}$ counts the number of
localized logical operators in $A$ while $g_{B}$ counts the number of
localized logical operators in $B$. Localized logical operators in
$M_{AB}$ are counted twice both in $g_{A}$ and $g_{B}$ while non-local
logical operators are not counted either in $g_{A}$ or
$g_{B}$. Localized logical operators in $M_{A}$ and $M_{B}$ are
counted once in $g_{A}$ and $g_{B}$ respectively.

Then, a simple number counting argument leads to the following
corollary.

\begin{corollary}
The number of independent localized logical operators in $M_{AB}$ and
the number of independent non-local logical operators are the same.
\end{corollary}

The proof is immediate with Theorem \ref{theorem1}. Let us denote the
number of independent logical operators in $M_{A}$, $M_{B}$, $M_{AB}$
and $M_{\phi}$ as $m_{A}$, $m_{B}$, $m_{AB}$ and $m_{\phi}$ where
\begin{align}
m_{AB}&= G(M_{AB}) - G(\mathcal{S})\\
m_{A}&=G(L_{A}) - G(M_{AB})\\
m_{B}&=G(L_{B}) - G(M_{AB})\\
m_{\phi} &= 2k - (m_{A} + m_{B} +m_{AB} ).
\end{align}
Since $g_{A} =m_{A} + m_{AB}$ and $g_{B}=m_{B}+m_{AB}$, theorem \ref{theorem1} asserts that
\begin{align}
m_{A} + m_{B} + 2 m_{AB}=2k.
\end{align}
Therefore, we have
\begin{align}
m_{AB}=m_{\phi}.
\end{align}

This corollary complements our framework by showing that non-local
logical operators $\delta_{i}$ in $\Delta$ are all the possible
independent non-local logical operators due to $m_{AB}=m_{\phi}$.

Second, let us discuss some consequences of Theorem \ref{theorem1} for
coding properties. Theorem \ref{theorem1} allows us to derive the
quantum singleton bound easily \cite{KL97}. Let us define three
subsets such that $V_{A}=d-1$, $B=\bar{A}$ and $B' \subseteq B$ with
$V_{B'}=d-1$. Then, we have
\begin{align}
g_{A}&=g_{B'}=0 \\
g_{B}&=2k.
\end{align}
Since we have
\begin{align}
g_{B} \leq 2(V_{B} -V_{B'} ),
\end{align}
we obtain
\begin{align}
k \leq N- 2(d-1).
\end{align}
This is the quantum singleton bound.

\subsubsection{Geometric interpretation of the theorem}

Now, let us apply Theorem~\ref{theorem1} to a stabilizer code which is
implemented with some physical qubits where qubits are located at some
specific positions in geometric manifolds. To begin with, let us
consider two regions $A$ and $A'\subset A$ which can support the same
number of independent logical operators with $g_{A}=g_{A'}$
(Fig.\ref{bipartition_theorem}(b)). Note that all the logical
operators defined inside $A$ can be transformed into equivalent
logical operators defined inside $A'$ by applying appropriate
stabilizers. Therefore, \textit{all the logical operators in $A$ can
  be deformed into other equivalent logical operators defined inside a
  smaller subset $A'$}.

Theorem~\ref{theorem1} clearly shows limitations on the possible
geometric shapes of equivalent logical operators. In order to create a
quantum code with large logical operators, we need to have a small
$g_{A}$ for large region $A$. However, the effort of decreasing
$g_{A}$ for large $A$ results in increasing $g_{B}$ for small $B$ for
$B=\bar{A}$ since we have $g_{A}+g_{B}=2k$. Thus, our theorem shows a
clear restriction on the \textit{sizes of logical operators}, and
indicates the \textit{intrinsic duality on geometric shapes of
  localized logical operators} of stabilizer codes in a bi-partition.

Now that we have the restriction $g_{A}+g_{B}=2k$ in hand, let us discuss
the problem of giving an upper bound on the code distance for a local
stabilizer code. This problem has been addressed in the literature
\cite{Local_Stabilizer}, by a beautiful construction of logical
operators in which logical operators can be shorten to equivalent
logical operators defined in smaller subsets. The \textit{cleaning
  lemma}, at the heart of this method, can now be understood as
resulting from an application of our formula. Let us suppose that
there is no logical operators defined inside $A$ at all. Then, one has
$g_{A}=0$ and $g_{B}=2k$ from Theorem \ref{theorem1}. Since $B$ can
contain $2k$ logical operators, we notice that \textit{all the logical
  operators in the system can be defined in $B$} by applying
appropriate stabilizers. Therefore, by finding a region $A$ such that
$g_{A}=0$, we can deform logical operators to its complement $A$. This
eventually gives an upper bound on the sizes of logical operators and
the code distances.

\subsubsection{Application to topological order}


Global symmetries of the system, which emerge as the existence of
large logical operators, are at the heart of topologically ordered
systems. To understand this underlying connection between global
symmetries and topological order, logical operators need to be
analyzed. For this purpose, our framework, and Theorem~\ref{theorem1},
give useful insights about the geometric properties of logical
operators.  We now apply these to analyze the geometric properties of
logical operators in the Toric code.

\begin{figure}[ht!]
\begin{center}
\includegraphics[width=0.95\linewidth]{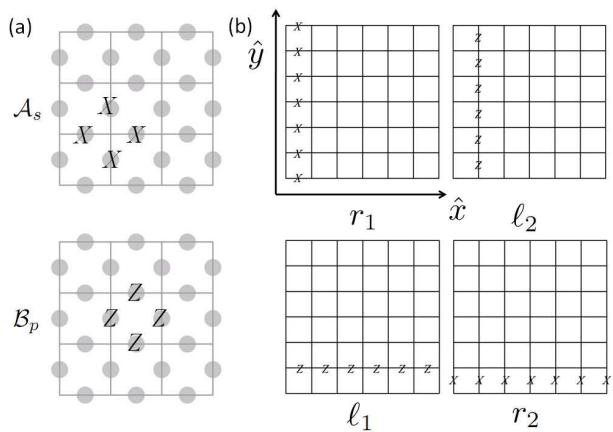}
\caption{The Toric Code. Qutbis live on edges on the bonds. Periodic boundary conditions are set in $x$
  and $y$ directions. (a) Stabilizers (interaction terms) in the Toric
  code. Stabilizers act on four qubits in either stars or
  plaquettes. (b) Logical operators in the Toric code.}
\label{Toric_Stabilizer}
\end{center}
\end{figure}

The Toric code is the simplest known, exactly solvable, model which is
described in the stabilizer formalism, supporting a degenerate ground
state, with topological order. Consider an $L \times L$ square lattice on the torus. The Toric code is defined qubits which live on edges of bonds
%
%
(Fig.\ref{Toric_Stabilizer} (a)). There are $N = 2 (L \times L)$ qubits
in total. For simplicity of discussion, we set periodic boundary
conditions. The Hamiltonian is:
\begin{align}
H &= - \sum_{s,p} (\mathcal{A}_{s} + \mathcal{B}_{p})\\
\mathcal{A}_{s} &= \prod_{i \in s} X_{i}\\
\mathcal{B}_{p} &= \prod_{i \in p} Z_{i}
\end{align}
where $s$ represent ``stars'' and $p$ represent ``plaquettes''
(Fig.\ref{Toric_Stabilizer}(a)). The Toric code has $2k =4$
independent logical operators since $G(\mathcal{S})=N-2$. Each of
independent logical operators $\ell_{1}$, $r_{1}$,
$\ell_{2}$ and $r_{2}$ are shown in
Fig.\ref{Toric_Stabilizer}(b). These four logical operators obey the
following commutation relations:
\begin{align}
\left\{
\begin{array}{cc}
  \ell_{1}  &, \ell_{2}  \\
  r_{1}  &, r_{2}               
\end{array}
 \right\}
\end{align}
where commutation relations are defined in a way similar to the canonical representation.

\begin{figure}[ht!]
\begin{center}
\includegraphics[width=0.9\linewidth]{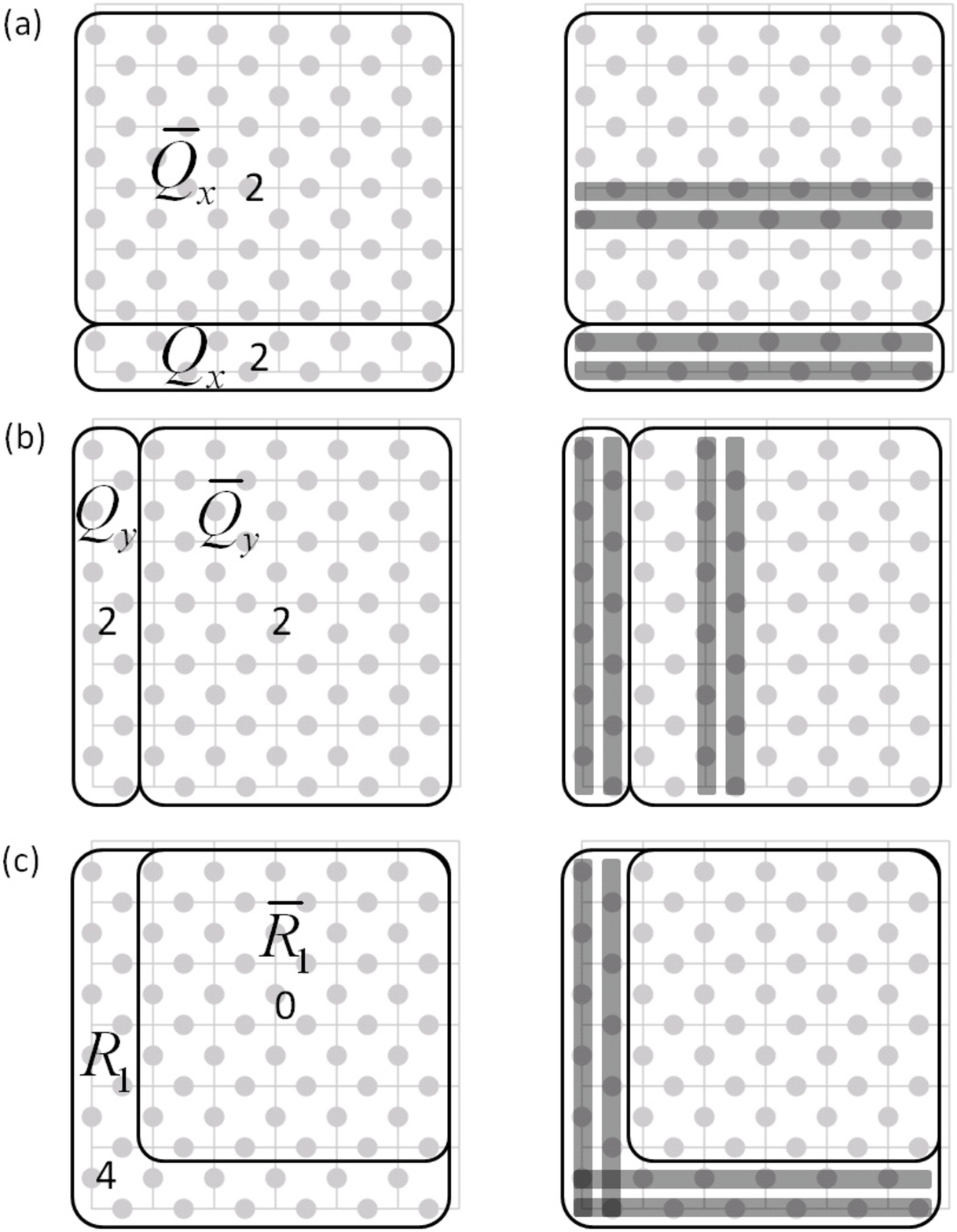}
\caption{Various bi-partitions and geometric properties of logical
  operators in the Toric code. Rectangles represent logical operators
  and circles represent qubits. Periodic boundary conditions are set
  in the $\hat{x}$ and $\hat{y}$ directions. Figures on the left hand side
  show bi-partitions and the number of logical operators defined
  inside each of complementary regions. Figures on the right hand side
  show the shapes of logical operators. (a) A bi-partition into
  $Q_{x}$ and $\bar{Q}_{x}$. Each dot represents each qubit. $Q_{x}$ is a region of qubits which extends in the $\hat{x}$
  direction.  Logical operators connected by two-sided arrows are
  equivalent. Both $Q_{x}$ and $\bar{Q}_{x}$ support only
  $r_{1}$ and $\ell_{2}$. $r_{2}$ and $\ell_{1}$
  cannot be defined wither inside $Q_{x}$ or $\bar{Q}_{x}$. (b) A
  bi-partition into $Q_{x}$ and $\bar{Q}_{x}$. Both $Q_{y}$ and
  $\bar{Q}_{y}$ support only $r_{2}$ and
  $\ell_{1}$. $r_{1}$ and $\ell_{2}$ cannot be defined
  wither inside $Q_{y}$ or $\bar{Q}_{y}$. (c) A bi-partition into
  $R_{1} = Q_{x}\cup Q_{y}$. $R_{1}$ supports all the independent
  logical operators $r_{1}$, $\ell_{1}$, $r_{2}$ and
  $\ell_{2}$. There is no logical operator defined inside
  $\bar{R}_{1}$.}
\label{Toric2}
\end{center}
\end{figure}

Consider several bi-partitions of the system and discuss how the
geometric properties of the logical operators may be understood. Let
us define the following regions. $Q_{x}$ is a region which circles the
lattice in $\hat{x}$-direction (Fig.\ref{Toric2}(a)). $Q_{y}$ is a region
which circles the lattice in the $\hat{y}$-direction
(Fig.\ref{Toric2}(b)). Finally, $R_{1}=Q_{x} \cup Q_{y}$ is a region
which is the union of $Q_{x}$ and $Q_{y}$ (Fig.\ref{Toric2}(c)).

Consider first a bi-partition into two subsets $Q_{x}$ and
$\bar{Q}_{x}$ described in Fig.\ref{Toric2}(a). Logical operators
$r_{2}$ and $\ell_{1}$ are defined inside $Q_{x}$, and we
have $g_{Q_{x}}\geq 2$. $r_{2}$ and $\ell_{1}$ also have
equivalent logical operators in $\bar{Q}_{x}$ since translations of
$r_{2}$ and $\ell_{1}$ in $\hat{y}$-direction are equivalent to
%
%
original logical operators $r_{2}$ and $\ell_{1}$
respectively. Then, we have $g_{\bar{Q}_{x}}\geq 2$.

Now apply theorem \ref{theorem1} to this bi-partition. Since
$g_{Q_{x}} + g_{\bar{Q}_{x}}=4$ due to the theorem, we must have
$g_{Q_{x}} = g_{\bar{Q}_{x}}=2$. Let us interpret these equations and
discuss the geometries of logical operators. Since $g_{\bar{Q}_{x}}
=2$, $r_{1}$ and $\ell_{2}$ are all the independent logical
operators which can be defined inside $Q_{x}$. In other words,
\textit{logical operators $r_{2}$ and $\ell_{1}$ cannot both
  be defined inside $Q_{x}$}. Without discussing the properties of
$r_{2}$ and $\ell_{1}$, one can analyze geometric properties
of $r_{2}$ and $\ell_{1}$ through Theorem
\ref{theorem1}. This observation can be explained through the
classifications of logical operators in our framework. Logical
operators $r_{1}$ and $\ell_{2}$ are localized logical
operators in a set $M_{Q_{x}\bar{Q}_{x}}$ while logical operators
$r_{2}$ and $\ell_{1}$ are non-local logical operators in a
set $M_{\phi}$ in a bi-partition into $Q_{x}$ and $\bar{Q}_{x}$.

It is more illuminating when we consider the equation $g_{\bar{Q}_{x}}
=2$. Even when we expand the region $Q_{x}$ to $\bar{Q}_{x}$, logical
operators $r_{2}$ and $\ell_{1}$ still cannot be defined
inside $\bar{Q}_{x}$ since $r_{1}$ and $\ell_{2}$ are all the
independent logical operators which can be defined inside
$\bar{Q}_{x}$. Therefore, one can conclude that logical operators
$r_{2}$ and $\ell_{1}$ can be defined \textit{only inside
  regions which circle around the lattice in the $\hat{y}$ direction}. The
similar discussion holds for logical operators $r_{1}$ and
$\ell_{2}$ by considering a bi-partition into $Q_{y}$ and
$\bar{Q}_{y}$ (Fig.\ref{Toric2}(b)). Logical operators $r_{1}$
and $\ell_{2}$ can be defined \textit{only inside regions which
  circle around the lattice in the $\hat{x}$ direction}.

Next, let us consider a bi-partition into two subsets $R_{1}=Q_{x}\cup
Q_{y}$ and $\bar{R}_{1}$ described in Fig.\ref{Toric2}(c). Since all
the four independent logical operators can be defined inside $R_{1}$,
we have $g_{R_{1}}=4$. Then, there is no logical operator which can be
defined inside $\bar{R}_{1}$ since $g_{\bar{R}_{1}}=0$. One notices
that $\bar{R}_{1}$ has no winding either in $x$ or $y$
direction. Therefore, \textit{there is no logical operator defined
  inside a region which does not circle around the lattice in any
  direction}.

These discussions clarify that the geometric shapes of logical
operators have universal, {\em topological} properties, which are
invariant under the application of stabilizers.  Specifically, whether
a region circles around the lattice in $x$ and $y$ directions can be
quantified by the \textit{winding numbers} of regions.  Define the
winding numbers $w_{i}$ such that $w_{i}=1$ if a region circles around
the lattice in $i$ direction where $i = x,y$ and $w_{i}$
otherwise. The winding numbers $w_{i}$ of geometric shapes of logical
operators are quantities which are invariant among all the equivalent
logical operators. Thus, \textit{the winding numbers of logical
  operators may be viewed as topological invariants}.

This analysis of the logical operators of the Toric code shows that
the signature of quantum order in the system can be found in the
geometric properties of symmetry operators which commute with the
system Hamiltonian. Though logical operators are originally used as
indicators for coding properties of quantum codes, logical operators
can actually be the indicators for quantum order, including
topological order. Our framework for logical operators, coupled with
appropriate geometric considerations, can be used to study such order,
and indeed may be useful for classifying quantum phases of such
systems.

\subsection{Non-Locality of Logical Qubits and Secret-Sharing \label{logical_qubits}}

%

Perhaps as interesting as the geometric properties of logical
operators is the nature of entanglement of degenerate ground
states. The entanglement of single stabilizer states may be
characterized by entanglement entropy \cite{FC04}. However, in
stabilizer codes, the nature of the entanglement in each of several
degenerate ground states can be different. This makes a stabilizer
code capable of providing a kind of entanglement which is essentially
different from the entanglement of a single state. Specifically, the
\textit{distributed nature of entanglement in ground state space} is
of interest.

The study of entanglement distributed across two or more parties is
enabled by the analysis of logical operators provided in our
framework. Below, using the framework, we discuss how entanglement is
distributed in a bi-partition by extending the classification of
logical operators to the classification of logical qubits. In
particular, we quantify how much entanglement is distributed across a
bi-partition and discuss secret-sharing of information \cite{secret1}
between two parties.

In stabilizer codes, degenerate ground states of the Hamiltonian are
the same as logical qubits of the corresponding stabilizer code. In
order to understand entanglement of degenerate ground states in a
stabilizer code, logical qubits need to be analyzed. Since logical
qubits are described by a pair of anti-commuting logical operators,
commutation relationships between all the classified logical
operators, including non-local logical operators, must be
understood. For this purpose, all the logical operators need to be
listed along with their classifications.

In section \ref{sec:classification}, we start our discussion by
analyzing the commutation relations between each of $2k$ logical
operators which are classified and computed through our framework. We
introduce a classification of logical qubits by extending the
classification of logical operators. In section \ref{sec:secrecy}, we
discuss secret-sharing of classical and quantum information and show
that our classification of logical qubits clearly capture distribution
of entanglement in a stabilizer code.

\subsubsection{Non-local logical qubit}
\label{sec:classification}

In order to discuss properties of logical qubits, properties of pairs
of anti-commuting logical operators need to be analyzed. Hence, let us
start by listing all the classified logical operators obtained through
our framework. Logical operators in $M_{A}$, $M_{B}$, $M_{AB}$ and
$M_{\phi}$ are found in the following group of operators:
\begin{align}
\mathcal{C}(\mathcal{S}_{A}) &=
\left\langle
\begin{array}{cccccc}
 \{ \bar{X}_{i} \} &, \mathcal{S}_{A} , & \ell_{1}, & \cdots ,& \ell_{d} ,& \{ \alpha_{i} \}   \\
 \{ \bar{Z}_{i} \} &, \phi            , & r_{1}, & \cdots ,& r_{d} ,& \{ \alpha_{i}' \}        
\end{array}
 \right\rangle
\end{align}
and
\begin{align}
\mathcal{C}(\mathcal{S}_{B}) &=
\left\langle
\begin{array}{cccccc}
 \{ \bar{X}_{i}' \} &, \mathcal{S}_{B} , & \ell_{1}', & \cdots ,& \ell_{d}' ,& \{ \beta_{i} \}   \\
 \{ \bar{Z}_{i}' \} &, \phi            , & r_{1}', & \cdots ,& r_{d}' ,& \{ \beta_{i}' \}        
\end{array}
 \right\rangle
\end{align}
where $\ell_{i}$ are in $M_{AB}$, $\alpha_{i}$ and $\alpha_{i}'$ are in $M_{A}$, $\beta_{i}$ and $\beta_{i}'$ are in $M_{B}$, and $\delta_{i}=r_{i}r_{i}'$ are in $M_{\phi}$. 

\begin{figure}[ht!]
\begin{center}
\includegraphics[width=1.0\linewidth]{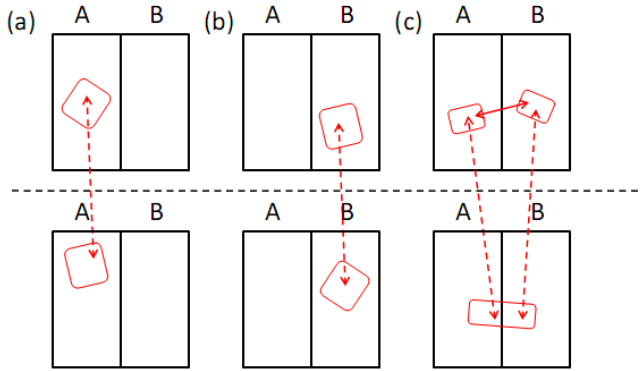}
\caption{The commutation relations between the four subsets of logical
  operators. The dotted arrows represent the anti-commutations between
  corresponding logical operators. We may have anti-commutations
  between two logical operators in the same column while other pairs
  always commute each other as in the symbolic form of canonical
  representations. (a) Anti-commutations inside $M_{A}$. (b)
  Anti-commutations inside $M_{B}$. (c) The pairing between $M_{AB}$
  and $M_{\phi}$. (a)(b) define local logical qubits while (c) defines
  non-local logical qubits.}
\label{commutation}
\end{center}
\end{figure}

Then, in order to specify commutation relationships between all the
classified logical operators, we list all the logical operators
symbolically as follows:
\begin{align}
\left\{
\begin{array}{ccccccccc}
  \alpha_{1}  &, \cdots , & \alpha_{m_{A}'} , & \beta_{1}  &, \cdots , & \beta_{m_{B}'} , & \ell_{1} ,&  \cdots ,& \ell_{m_{\phi}}   \\
  \alpha_{1}' &, \cdots  ,  & \alpha_{m_{A}'}' , & \beta_{1}'  &, \cdots , & \beta_{m_{B}'}' ,& \delta_{1} ,&    \cdots ,& \delta_{m_{\phi}}'        
\end{array}
 \right\} \label{choice}
\end{align}
where two logical operators in the same column anti-commute each other
as in the canonical representation and $m_{A}' = \frac{1}{2}m_{A}$ and
$m_{B}' = \frac{1}{2}m_{B}$. We write down the commutation relations
between logical operators which describe logical qubits symbolically
as follows
\begin{equation}
\begin{split}
M_{A} &\leftrightarrow  M_{A} \\
M_{B} &\leftrightarrow  M_{B} \\
M_{\phi} &\leftrightarrow M_{AB} \label{commutation_eq}
\end{split}
\end{equation}
where the two-directional arrow "$\leftrightarrow$" represents the
possibility of anti-commutations between corresponding sets of logical
operators. We summarize the above commutation relationships
graphically in Fig.\ref{commutation}.

This list of logical operators in Eq.(\ref{choice}) defines $k$
logical qubits. Let us analyze the locality and non-locality of these
logical qubits in detail. First, consider a logical qubit described by
a pair of anti-commuting logical operators $\alpha_{i}$ and
$\alpha_{i}'$ (Fig.\ref{commutation}(a)). This logical qubit is
described by a pair of two localized logical operators. There also
exist logical qubits described by a pair of localized logical
operators in $M_{B}$ (Fig.\ref{commutation}(b)). We call these logical
qubits \textit{local logical qubits}. Such local logical qubits can be
completely manipulated through local operations on physical qubits
inside either $A$ or $B$.

Next, let us consider a logical qubit described by a pair of
anti-commuting logical operators $\ell_{i}$ and $\delta_{i}$
(Fig.\ref{commutation}(c)). This logical qubit is described by one
localized logical operator and one non-local logical operator. We call
such a logical qubit \textit{local logical qubit}. Non-local logical
qubits cannot be completely manipulated through local operations on
physical qubits either inside $A$ or inside $B$. Therefore, $k$
logical qubits defined in Eq.(\ref{choice}) can be classified into two
types, local and non-local logical qubits.

Recall the problem of the arbitrariness in the definition of logical
qubits which we mentioned at the end of section
\ref{sec:setups}. Though the above choice of anti-commuting logical
operators defines $k$ different logical qubits, one can choose a
different set of logical operators to define logical qubits. However,
we shall see that the above definition of logical qubits and
classification into local and non-local logical qubits are quite
useful when distributions of entanglement are to be analyzed.

\subsubsection{Secret-sharing}
\label{sec:secrecy}

Let us now analyze how secret-charing schemes work with a stabilizer
code. We also explain our results in terms of local and non-local
logical qubits.

Secret-sharing is a scheme which allows sharing of information between
two parties or multiple parties so that each party cannot access the
encoded information individually. Shared information can be accessed
only when all the parties agree and execute a protocol together. Such
encoded information is shared by multiple parties as a secret among
all the parties. Some entangled quantum system can be used as a
resource for secret-charing of classical or quantum information
\cite{secret1, secret2}. Below, we consider secret-sharing between two
parties with a stabilizer code.

First, we discuss secret-sharing of classical information between two
parties using a stabilizer code. Consider a situation where one party
$A$ possesses a subset of qubits $A$ and the other party $B$ possesses
a complementary subset of qubits $B$. When $\ell$ is a non-local logical
operator with $\ell \in M_{\phi}$, a bit of information can be shared
between $A$ and $B$ by assigning $0$ and $1$ to each of eigenstates of
$\ell$. Such encoded information cannot be read out by individual access
either from $A$ or from $B$ since $\ell$ does not have an equivalent
logical operator defined either inside $A$ or inside $B$. Therefore, a
bit of information can be shared between $A$ and $B$ only if there
exists a non-local logical operators with $m_{\phi}\not=0$.

The method for sharing one bit of information can be easily extended
to the method for sharing multiple bits of information. In order to
share $m$ bits of information between $A$ and $B$, there must be $m$
independent non-local logical operators. Suppose that $m$ bits of
information are encoded with respect to $u_{1}, \cdots , u_{m} \in
M_{\phi}$. If $m > m_{\phi}$, there exists a set of integers $R$ such
that
\begin{align}
\prod_{i \in R} u_{i} \in L_{A}\cup L_{B}.
\end{align}
Then, a local measurement can read out a bit of information out of
shared at most $m$ bits of information. Therefore, only $m_{\phi}$
bits of information can be shared between $A$ and $B$.

Let us interpret the above discussion in terms of logical qubits
defined in Eq.(\ref{choice}). A local logical qubit described by a
pair of logical operators $\alpha_{i}$ and $\alpha_{i}'$ or
$\beta_{i}$ and $\beta_{i}'$ cannot be used for sharing classical
information. A non-local logical qubit described by a pair of
$\delta_{i}$ and $\ell_{i}$ can be used for sharing a bit of
information. The classification of logical qubits introduced in
section \ref{sec:classification} directly corresponds to their
abilities as resources for sharing classical information.

Next, let us discuss secret-sharing of quantum information. In order
to encode quantum information, a pair of anti-commuting logical
operators $\ell$ and $r$ is required. Suppose that a qubit is encoded
with respect to $\ell$ and $r$.  In order for this encoded qubit to be
shared by $A$ and $B$, the following condition is necessary:
\begin{align}
\ell, r \in M_{\phi} \label{lr}.
\end{align}
However, this condition is not sufficient to provide a logical qubit
for sharing a qubit. If there exists a localized logical operator $\ell'
\in L_{A}\cup L_{B}$ which satisfies $\{\ell',\ell\}=0$ and $\{\ell',r\}=0$, a
measurement of $\ell'$ can have the same effect as a measurement of $\ell r$
on encoded qubit information. Similarly, a measurement of $\ell' \in
L_{A}\cup L_{B}$ with $\{\ell',\ell\}=0$ and $[\ell',r]=0$ have the same effect
as a measurement of $r$ on encoded qubit information and a measurement
of $\ell' \in L_{A}\cup L_{B}$ with $[\ell',\ell]=0$ and $\{\ell',r\}=0$ have the
same effect as a measurement of $\ell$ on encoded qubit
information. Therefore, necessary conditions for a logical qubit
described by $\ell$ and $r$ to be shared between $A$ and $B$ are:
\begin{equation}
\begin{split}
[\ell, \ell']&=0 \\
[r, \ell']&=0 \\
\{\ell,r\}&=0  \label{eq:condition}
\end{split}
\end{equation}
for all the logical operators $\ell'\in L_{A} \cup L_{B}$, along with the
condition in Eq.(\ref{lr}).

Now, let us show that there cannot exist a pair of logical operators
$\ell$ and $r$ which satisfies the above condition in
Eq.(\ref{eq:condition}). All the logical operators can be represented
as a product of $2k$ logical operators $\alpha_{i}$, $\alpha_{i}'$,
$\beta_{i}$, $\beta_{i}'$, $\ell_{i}$ and $\delta_{i}$ except the trivial
contribution from $\mathcal{S}$. Then $\ell$ and $r$ must be a product of
$\ell_{i}$ in order for $\ell$ and $r$ to commute with all the localized
logical operators. This contradicts the fact that $\ell$ and $r$ are
logical operators in $M_{\phi}$. Therefore, secret-sharing of quantum
information is impossible the bi-partitioning of a stabilizer
code. Our result can be summarized in the following theorem.

\begin{theorem}
One cannot share quantum information secretly between two parties
inside the ground state space of stabilizer codes.
\end{theorem}

Let us interpret this observation with respect to the classification
of logical qubits defined for the choices of logical operators in
Eq.(\ref{choice}). In order for quantum information to be shared
between $A$ and $B$, there should exist a logical qubit described by a
pair of non-local logical operators. However, there are only two types
of logical qubits, local logical qubits and non-local logical
qubits. There cannot exist a logical qubit which can be used for
sharing quantum information. We note that this is a direct consequence
of $m_{\phi}=m_{AB}$.

We have seen that non-local logical operators in non-local logical
qubits are essential in secret-sharing of classical information. It
might be natural to think that non-local logical operators are
responsible for sharing information in a non-local way between two
separated parties. However, a surprising consequence of the equation
$m_{\phi}=m_{AB}$ in secret-sharing of quantum information is the fact
that the properties of non-local logical operators are completely
governed by localized logical operators in $M_{AB}$. In fact, a
trivial, but very insightful interpretation of $m_{\phi}=m_{AB}$ in
terms of secret-sharing of classical information is obtained:

\begin{corollary}\label{criteria}
The necessary and sufficient condition for a stabilizer code to be
useful for classical information sharing is $m_{AB}\not =0$.
\end{corollary}

This corollary states the following: The existence of localized
logical operators in $M_{AB}$ automatically guarantees the existence
of non-local logical operators and non-local logical
qubits. Therefore, the existence of localized logical operators in
$M_{AB}$ can be used as a criteria to check whether a stabilizer code
and a given bi-partition can give a resource for secret-charing of
classical information.

\subsection{Entanglement Entropy}

As our third, and final example of application of our framework, we turn
to {\em entanglement entropy}.  Entanglement entropy plays important
roles in condensed matter physics and quantum information science
\cite{Entanglement_Criticality, Entropy_CFT_2D, Entanglement_QPT1,
  Li08, Topo1, Topo2}.  We may analyze the entanglement entropies of
states inside the ground state space, and obtain some nice bounds
using our framework.

For simplicity, we compute entanglement entropies for two ground
states $|\psi_{0}\rangle$ and $|\psi_{1}\rangle$ which satisfy
\begin{align}
\ell_{i}|\psi_{0}\rangle &= |\psi_{0}\rangle \\
r_{i}|\psi_{1}\rangle &= |\psi_{1}\rangle 
\end{align}
for $\forall i$.  Therefore, $|\psi_{0}\rangle$ is characterized by
logical operators $\ell_{i}$ in $M_{AB}$ while $|\psi_{1}\rangle$ is
characterized by logical operators $r_{i}$ in $M_{\phi}$.  Since
logical operators in $M_{A}$ and $M_{B}$ do not affect the
entanglement over $A$ and $B$, without loss of generality, we can
assume that
\begin{align}
f_{i}|\psi_{0}\rangle &= |\psi_{0}\rangle \\
f_{i}|\psi_{1}\rangle &= |\psi_{1}\rangle
\end{align}
where $f_{i}$ are $m_{A}' + m_{B}'$ independent logical operators in $M_{A}$ and $M_{B}$.
Then, we can represent $|\psi_{0}\rangle$ and $|\psi_{1}\rangle$ as
\begin{align}
|\psi_{0}\rangle \langle \psi_{0}| &= \frac{1}{2^{N}} \prod_{i=1}^{m_{\phi}}(I+\ell_{i}) \prod_{i=1}^{m_{A}'+m_{B}'}(I+ f_{i} ) \prod_{i=1}^{N-k}(I+S_{i}) \\
|\psi_{1}\rangle \langle \psi_{1}| &= \frac{1}{2^{N}} \prod_{i=1}^{m_{\phi}}(I+r_{i}) \prod_{i=1}^{m_{A}'+m_{B}'}(I+ f_{i} ) \prod_{i=1}^{N-k}(I+S_{i})
\end{align}
with a set of $N$ independent commuting Pauli operators which consist
of $N-k$ independent stabilizer generators $S_{i}$ and logical operators.  In order to compute
the entanglement entropy of $|\psi_{0}\rangle$ and $|\psi_{1}\rangle$,
we consider the following groups of operators
\begin{align}
\mathcal{S}(0) &= \langle \{\ell_{i}\} , \{f_{i}\} , \{S_{i}\}, \forall i \rangle \\
\mathcal{S}(1) &= \langle \{r_{i}\} , \{f_{i}\} , \{S_{i}\}, \forall i \rangle
\end{align}
with the decompositions 
\begin{align}
\mathcal{S}(0)&= \langle \mathcal{S}(0)_{A} , \mathcal{S}(0)_{B} , \mathcal{S}(0)_{AB} \rangle\\
\mathcal{S}(1)&= \langle \mathcal{S}(1)_{A} , \mathcal{S}(1)_{B} , \mathcal{S}(1)_{AB} \rangle
\end{align}
in a way similar to the decomposition of $\mathcal{S}$ into
$\mathcal{S} = \langle \mathcal{S}_{A}, \mathcal{S}_{B},
\mathcal{S}_{AB}\rangle$.  Then, the entanglement entropies of
$|\psi_{0}\rangle$ and $|\psi_{1}\rangle$ are represented as follows \cite{FC04}:
\begin{align}
E_{A}(|\psi_{0}\rangle) &= \frac{1}{2} G(\mathcal{S}(0)_{AB})  \\
E_{B}(|\psi_{1}\rangle) &= \frac{1}{2} G(\mathcal{S}(1)_{AB}).
\end{align}
Since $\mathcal{S}(0)$ includes
$\ell_{i}$, generators $\ell_{i}\ell_{i}'$ for $\mathcal{S}_{AB}$ can be
decomposed into $\ell_{i}$ and $\ell_{i}'$.  Therefore, we have
\begin{align}
E_{A}(|\psi_{0}) &= \frac{1}{2} G(\mathcal{S}(0)_{AB})  \\
                 &= \frac{1}{2} (G(\mathcal{S}_{AB}) - m_{\phi}).
\end{align}
On the other hand, since $\mathcal{S}(1)_{AB}$ includes $r_{i}$ in
addition to the generators for $\mathcal{S}_{AB}$, we have
\begin{align}
E_{A}(|\psi_{1}\rangle) = \frac{1}{2} (G(\mathcal{S}_{AB}) + m_{\phi}).
\end{align}

Thus, one can easily see that the entanglement entropy of an arbitrary
ground state $|\psi\rangle$ in the degenerate ground state space satisfies
\begin{align}
\frac{1}{2} (G(\mathcal{S}_{AB}) - m_{\phi}) \leq E_{A}(|\psi\rangle)
\leq \frac{1}{2} (G(\mathcal{S}_{AB}) + m_{\phi}).
\end{align}
Therefore, the entanglement entropy of degenerate ground states has an arbitrariness $m_{\phi}$, which is the same as the number of non-local logical operators. 

Now, let us apply this bound on the entanglement entropy to two problems we have addressed in this section. We begin with the relation between this bound and secret-sharing. Since the maximal entropy of degenerate ground states is $\frac{1}{2} (G(\mathcal{S}_{AB}) + m_{\phi})$ while the minimal entropy is $\frac{1}{2} (G(\mathcal{S}_{AB}) - m_{\phi}) $, the entropy may vary up to $m_{\phi}$ inside the degenerate ground state space. Therefore, the ground state space of the Hamiltonian is capable of encoding $m_{\phi}$ bits of information.  This is another interpretation of secret-sharing of classical information with logical qubits described by a pair of logical operators in $M_{\phi}$ and $M_{AB}$. Thus, the distribution of entanglement can be accessed in terms of the entanglement entropy and its relation with non-local logical operators in $M_{\phi}$.

Next, let us consider the entanglement entropy of degenerate ground states of the Toric code. The entanglement entropy for a region without windings is particularly important in the discussion of the entanglement area law \cite{Area_Law}. From the discussion on the Toric code in section \ref{geometry}, there is no logical operator inside a region without any winding. Then, one can tightly bound the entanglement entropy since $m_{\phi}=0$:
\begin{align}
E_{A}(|\psi\rangle) = \frac{1}{2} G(\mathcal{S}_{AB}).
\end{align}
Through a direct computation, we have
\begin{align}
E_{A} = 2(n_{x} + n_{y}) -2
\end{align}
where $A$ is an $n_{x} \times n_{y}$ square region of the lattice, with $2n_{x} \times n_{y}$ qubits. Since there are no logical operators inside $A$ for $n_{x},n_{y}<L$ where $L$ is the length of the entire lattice, the entanglement entropy takes a single value. Noticing that $2(n_{x} + n_{y})$ is the length of the perimeter of $A$, the term independent of length, the factor of two can be identified as being the topological entropy of the Toric code \cite{Topo1, Topo2}. 

Unlike secret-sharing, the entanglement entropy of the Toric code takes only a single value. This is a consequence of the fact that degenerate ground states of the Toric code are robust against local perturbations. In the language of logical operators, the robustness of ground states results from the fact that there are no logical operators inside local regions. Since all the logical operators are defined globally, any local perturbation cannot couple different ground states, so ground states are stable. In our discussion of the Toric code, we found that there is no logical operator inside regions without windings. This is the underlying reason behind the robustness of the Toric code and the resulting uniqueness of the entanglement entropy. Thus, the uniqueness of the entanglement entropy reflects the robustness of the Toric code, demonstrating the usefulness of analyzing the locality and non-locality of logical operators.

\section{Summary and Outlook}

In this paper, we have provided a systematic framework to study the
non-local properties of logical operators and ground states in the
stabilizer formalism, given a bi-partition. It is our hope that this
will open the door to further unite the studies of correlations in
condensed matter physics with the studies of entanglement in quantum
information science.  The framework can likely be broadened in many
ways, five of which are discussed below.

Though we have studied only bi-partite systems in this paper,
many-qubit systems can provide rich varieties of entanglement that may
not be quantified by a bi-partition. Even for three qubit states,
there are two inequivalent states, the W state and the GHZ state,
which can be classified only by multi-partite entanglement
\cite{three_qubit}. Several multi-partite entanglement measures have
been proposed to characterize many-qubit entangled states, and each of
them successfully reveals different aspects of quantum entanglement
\cite{n_tangle, 3_concurrence}. Our framework can be also generalized
to the study of stabilizer codes with a multi-partition by extending
the classification of logical operators based on their localities with
respect to each of subsets constituting a multi-partition.

Multi-partite correlations are also important in the study of
condensed matter physics, as seen in the investigation of spin
systems. One interesting example in which multi-partition plays a
crucial role is the study of topological order. In fact, the study of
topological order lies at the interface between condensed matter
physics and quantum information theory. However, currently,
topological order is interpreted in somewhat different ways in
condensed matter physics and quantum coding theory.

When topological order is discussed in quantum codes, properties of a
single state are frequently studied, rather than properties of the
whole system. When a quantum code possesses topological order, it is
known that there usually exist large logical operators that are
defined non-locally in a multi-partition. The existence of such large
logical operator often results in a degenerate ground state with
topological order \cite{Toric, Kitaev_1dim}. Topological properties of
a state that corresponds to a codeword in a quantum code can be
studied through its topological entropy \cite{Topo1, Topo2}. It has
also been suggested that the preparation time of a corresponding state
from a product state through local operations can be used to
distinguish globally entangled states \cite{Lieb}.

On the other hand, when topological order is discussed in condensed
matter physics, the entire spectrum of the Hamiltonian is discussed,
rather than a single state. Topological order often appears along with
small ground state degeneracy \cite{Wen_Text}. This degeneracy can
usually explained by the existence of symmetry operators that commute
with the entire Hamiltonian. Topological orders can be analyzed by
scattering of anyonic excitations \cite{Quantum_Double}.

Recently, the connection between these two different views of
topological order in condensed matter physics and quantum coding
theory has begun to be understood. The relation between anyonic
excitations and topological entropy is discussed \cite{Topo1,
  Topo2}. The existence of degenerate ground states can be understood
through logical operators and symmetry operators which commute with
the entire Hamiltonian. Also, it is suggested that the entanglement
spectrum of a ground state of the Hamiltonian may reveal the
properties of the whole energy spectrum \cite{Li08}. A study of
stabilizer codes in a multi-partition will provide further insights on
this connection since entanglement measures and logical operators can
be computed efficiently with our framework.

Hand in hand with the global symmetries of topological order are
local, physical symmetries of the system, such as translation
symmetries and geometric symmetries. These symmetries can limit the
properties of quantum codes, but also may simplify their
analysis. Quantum codes with constrained global symmetries are
particularly important and need to be studied since these symmetries
may reflect the nature of real physical systems. These additional
constraints translate into constraints on the structure of the
overlapping operator groups in our framework. For example, translation
symmetries of the system result in translation symmetries of the
overlapping operator groups. Therefore, these additional constraints
make it easier to implement our framework.

An example of the situation when realistic physical systems become
desirable is the study of self-correcting quantum memories. A
self-correcting memory is an ideal memory device which corrects errors
by itself through thermal dissipations. There exists a theoretical
proposal of a self-correcting quantum memory in four-dimensional space
\cite{4dim_Toric}. However, it was proven that there does not exist a
self-correcting quantum memory within the stabilizer formalism
\cite{Local_Stabilizer}. The feasibility of three-dimensional
self-correcting quantum memory is an important open question which
remains unsolved at this moment. The memory's feasibility is related
to whether random thermal noises can accidentally create one of
logical operators which describe a logical qubit or not. The
probability for thermal noise to destroy encoded information can be
estimated by analyzing the subsets where each of logical operators is
defined. Our framework can study whether a logical operator can be
defined inside a subset or not. Therefore, it provides a useful tool
to investigate this interesting open question \cite{Beni10}.

Though we have limited our considerations to stabilizer codes, there are
many quantum codes and spin systems which cannot be described through the
stabilizer formalism. One novel class of quantum codes, now called {\em
subsystem codes} \cite{Shor95, Doucot05}, replaces the commuting interaction terms of
stabilizer codes (Eq.(\ref{stabHam})) with interaction terms which may
anti-commute with each other.  These subsystem codes are interesting,
particularly because they may potentially provide a means to realize a
self-correcting quantum memory in three-dimensions, for example, using the
Bacon-Shor subsystem code \cite{Bacon_Shor}. The two-dimensional Bacon-Shor
subsystem code also possess several promising features, such as a lower
fault-tolerance threshold \cite{Bacon_Shor_Thresh} compared with certain
codes utilizing similar space and time resources.  And in the condensed
matter physics community, the Bacon-Shor subsystem code is known as the
quantum compass model \cite{Kugel73}. Physical properties arising from the
quantum compass model have been studied
numerically \cite{Dorier05,Chen_b07,Orus09}, and are interesting, for
example, in the notable role they play in explaining the effects of the
orbital degree of freedom of atomic electrons on the properties of
transition-metal oxides \cite{Tokura00}.

Despite this promising progress with subsystem codes, the underlying
mechanism for the general capabilities of and physical properties arising
from stabilizer codes is not fully understood.  In subsystem codes, the
logical operators can be computed by analyzing all the elements in the
Hamiltonian, but this is generally computationally difficult.  The canonical
representations of our framework could allow efficient computation of
logical operators in subsystem codes by allowing the extraction of commuting
operators from the elements in the Hamiltonian.  Also, the analysis of
logical operators through our framework may be extended to subsystem codes
and give insights on the underlying mechanism of subsystem codes.

These subsystem codes also have interesting interpretations in terms
of condensed matter physics.  The Hamiltonian corresponding to a
subsystem code may contain interaction terms that anti-commute with
each other.  Unlike a Hamiltonian in the stabilizer formalism, a
ground state of such a Hamiltonian cannot be obtained by separately
minimizing each interaction term.  This is an analogue of frustration
in condensed matter physics.  A frustrated Hamiltonian with
anti-commuting interaction terms is not solvable in general.  However, by extracting the operator elements that commute with all the
interaction terms in the Hamiltonian, one can obtain some insight
about the physical properties appearing in the Hamiltonian even
without solving it.  These symmetry operators can be computed in a way
similar to the method employed in our framework for the computation of
logical operators through canonical representations.



\end{document}